# Contrasting H-mode behaviour with deuterium fuelling and nitrogen seeding in the all-carbon and metallic versions of JET

G. P. Maddison [1], C. Giroud [1], B. Alper [1], G. Arnoux [1], I. Balboa [1], M. N. A. Beurskens [1], A. Boboc [1], S. Brezinsek [2], M. Brix [1], M. Clever [2], R. Coelho [3], J. W. Coenen [2], I. Coffey [4], P. C. da Silva Aresta Belo [3], S. Devaux [5], P. Devynck [6], T. Eich [5], R. C. Felton [1], J. Flanagan [1], L. Frassinetti [7], L. Garzotti [1], M. Groth [8], S. Jachmich [9], A. Järvinen [8], E. Joffrin [6], M. A. H. Kempenaars [1], U. Kruezi [2], K. D. Lawson [1], M. Lehnen [2], M. J. Leyland [10], Y. Liu [11], P. J. Lomas [1], C. G. Lowry [12], S. Marsen [5], G. F. Matthews [1], G. K. McCormick [5], A. G. Meigs [1], A. W. Morris [1], R. Neu [5], I. M. Nunes [3], M. Oberkofler [5], F. G. Rimini [1], S. Saarelma [1], B. Sieglin [5], A. C. C. Sips [12], A. Sirinelli [1], M. F. Stamp [1], G. J. van Rooij [13], D. J. Ward [1], M. Wischmeier [5]
and JET EFDA contributors [†]

JET-EFDA, Culham Science Centre, Abingdon, OX14 3DB, UK.

[1] *EURATOM/CCFE Fusion Association, Culham Science Centre, Abingdon, Oxon. OX14 3DB, UK.*
[2] *IEK-Plasmaphysik, Forschungszentrum Jülich, Association EURATOM-FZJ, Jülich, Germany.*
[3] *IPFN, EURATOM-IST Associação, 1096 Lisbon, Portugal.*
[4] *Queen's University, Belfast BT7 1NN, Northern Ireland, UK.*
[5] *Max-Planck-Institut für Plasmaphysik, EURATOM-Association, 85748 Garching, Germany.*
[6] *CEA-Cadarache, Association Euratom-CEA, 13108 St Paul-lez-Durance, France.*
[7] *Division of Fusion Plasma Physics, Association EURATOM-VR , KTH, SE-10044 Stockholm, Sweden.*
[8] *Aalto University, Association EURATOM-Tekes, Otakaari 4, 02150 Espoo, Finland.*
[9] *Association Euratom-Etat Belge, ERM-KMS, Brussels, Belgium.*
[10] *York Plasma Institute, University of York, Heslington, York YO10 5DD, UK.*
[11] *Institute of Plasma Physics, Chinese Academy of Sciences, Hefei 230031, China.*
[12] *JET-EFDA/CSU, Culham Science Centre, Abingdon, Oxon. OX14 3DB, UK.*
[13] *Association EURATOM-FOM, PO BOX 1207, 3430 BE Nieuwegein, Netherlands.*

---

[†] *see appendix of F. Romanelli et al, Proceedings of the 24th IAEA Fusion Energy Conference, October 2012, San Diego, USA.*

**Abstract**

The former all-carbon wall on JET has been replaced with beryllium in the main torus and tungsten in the divertor to mimic the surface materials envisaged for ITER. Comparisons are presented between Type I H-mode characteristics in each design by examining respective scans over deuterium fuelling and impurity seeding, required to ameliorate exhaust loads both in JET at full capability and in ITER. Attention is focused upon a common high-triangularity, single-null divertor configuration at 2.5 MA, $q_{95} \approx 3.5$ yielding the most robust all-C performance. Contrasting results between the alternative linings are found firstly in unseeded plasmas, for which purity is improved and intrinsic radiation reduced in the ITER-like wall (ILW) but normalised energy confinement is $\approx 30\%$ lower than in all-C counterparts, owing to a commensurately lower (electron) pedestal temperature. Divertor recycling is also radically altered, with slower, inboard-outboard asymmetric transients at ELMs and spontaneous oscillations in between them. Secondly, nitrogen seeding elicits opposite responses in the ILW to all-C experience, tending to raise plasma density, reduce ELM frequency, and above all to recover (electron) pedestal pressure, hence global confinement, almost back to previous levels. A hitherto unrecognised role of light impurities in pedestal stability and dynamics is consequently suggested. Thirdly, while heat loads on the divertor outboard target between ELMs are successfully reduced in proportion to the radiative cooling and ELM recurrence effects of N in both wall environments, more surprisingly, average power ejected by ELMs and estimated scrape-off-layer pressure ratio during the approach to outboard detachment also both decline in the same proportion for the ILW. Finally, inter-ELM W sources in the ILW divertor tend to fall with N input, although core accumulation possibly due to net ELM erosion still leads to significantly less steady conditions than in all-C plasmas. This limitation of ILW H-modes so far will be readdressed in future campaigns to continue progress towards a fully-integrated scenario suitable for D-T experiments on JET and for "baseline" operation on ITER. The diverse changes in behaviour between all-C and ILW contexts demonstrate essentially the strong impact which boundary conditions and intrinsic impurities can have on tokamak-plasma states.

**Contents**

# 1. Introduction

It was recognised very early in the development of magnetic fusion that a major challenge would be controlled exhaust of plasma particles and power from a large tokamak device[1-5]. To address this issue, while helping to direct plasma interactions with material boundaries towards dedicated surfaces, the magnetic divertor[2,3,5-7] was introduced, focusing particularly on the poloidal concept[5-8] both to economise on extra external fields required and to enlarge divertor volume through axisymmetry. Design of first-generation burning devices like ITER[9-12] and power reactors like DEMO[13-16] still showed, however, that gross efflux power would exceed the handling capability of their divertor targets, requiring a large fraction of it instead to be dispersed over greater wall areas by induced radiation[2,17,18]. In fact, a radiated power fraction of $\approx 85$ - $90\%$ will probably be required during steady operation of these machines, which would be achieved mainly through line emissions from added impurity species[12,14]. Experimental investigation of the effects of seeding next-step-related plasmas with extrinsic impurities has therefore become a topic of mounting importance on many tokamaks over recent years[19-51].

At the same time, initial trials of deuterium-tritium mixtures particularly in the JET machine revealed a tendency for fuel isotopes to be retained within the torus by trapping in eroded and redeposited layers of the then preferred plasma-facing substance, carbon[52-54]. This mechanism could rapidly accumulate an inventory at the licensing limit for a project like ITER[10,12], so that in the absence of simple removal techniques, use of carbon became deprecated. Alternative metal surfaces expected to be less liable to co-deposition were therefore proposed, even though at the time, no existing device incorporated the combination of beryllium main-torus and tungsten divertor linings envisaged[9,10,12,55-57]. In response, a major refit of JET was undertaken to install exactly this distribution of novel materials and so to construct a unique ITER-like wall (ILW)[57-61]. Experiments exploring the implications for and impact on plasma operation have started from late 2011[62-65].

Installation of the ILW consequently presented an unparalleled opportunity to compare plasma behaviour in the previous all-carbon and present all-metal surface environments. Moreover, in JET such comparisons can be made for combinations of dimensionless parameters (normalised Larmor radius, pressure, collisionality, Greenwald fraction[66] etc.) most closely approaching those of ITER for any working tokamak[67,68]. Extrinsic seeding to raise radiation fraction clearly remains a key area of interest, and in fact it assumes a high priority specifically for the ILW itself, since a similar need to moderate divertor heat loads is expected to apply in JET as its auxiliary heating capacity is progressively raised up to $\approx 34$ MW and its pulse flat-top duration extended to $\approx 20$ s in forthcoming campaigns[69]. One consideration, for example, is that sensitivity to injection of impurities may be exacerbated in a tungsten divertor owing to their greater sputtering power than fuel species on metal targets[70,71], contamination by which erosion the plasma is then much less able to tolerate[72]. In effect, a state must be maintained where cooling by added impurities dominates over their tendency to admit heavier intrinsic particles. Systematic scans of deuterium fuelling and extrinsic seeding have therefore been conducted in the ILW, covering similar ranges and reproducing as closely as possible the operational parameters of counterpart scans previously performed in the all-C version of JET[42,43,44] (see Fig. 1). As in these predecessors, the goal is thus to span variations wide enough to be able to interpolate, rather than trying to extrapolate, to optimum conditions. Attention has been focused first on Type I ELMy H-mode, as the scenario foreseen for "baseline" inductive plasmas in ITER[11,12], and initially on nitrogen as the seed impurity owing to its suitability as a radiator at typical JET divertor temperatures[73], which favours a trend towards higher divertor over main-plasma emission, as seen in the all-C cases[44].

In conjunction with impurity seeding, adjustment of majority fuelling is suggested since higher divertor recycling is itself able to lower local plasma temperatures and to maximise volumetric power losses through increased density and target ion flux[8,74]. Eventually this can develop into the detached regime, where volumetric losses dominate over then suppress plasma fluxes to the targets[75], ostensibly defining an ideal power exhaust scheme for ITER[12] or DEMO[14]. Combination with extrinsic impurities remains necessary, however, since a pure fuelling approach to detachment at high power levels has too adverse an effect on global confinement[44,75], so that as hinted above an optimum balance, and even possible synergy[24], between the two inputs is required to obtain the best reconciliation of tolerable divertor loads with plasma performance. Importantly it was also found in all-C JET studies that energy confinement rises with higher magnetic shaping[76,77] and most notably that its robustness to increasing fuelling was much improved above a certain degree of triangularity, giving access to a new well-confined, high-density regime of lower ELM frequency and enhanced inter-ELM effluxes[77-79]. Consequently, scans in both all-C and ILW environments have been conducted first in a high-triangularity geometry (see Fig. 1) at a current of 2.5 MA and edge safety factor of $q_{95} \approx 3.5$, for which parameters the resilience to gas puffing was most pronounced[80]. The latter $q_{95}$ value further helped to reduce the occurrence of deleterious neo-classical tearing modes (NTMs) too[42,43].

The most immediate effect of replacing carbon with metal surfaces is likely to be on recycling behaviour, in terms both of particles and energy, with the former (C) tending to be more adsorbing and the latter (Be/W) more reflecting not only for fuel species, but also for nitrogen[81-84]. Hence the distribution of dominant sources and sinks, in both configuration and velocity space, around the plasma periphery may be significantly altered even when inputs and other principal paremeters are well matched. Such a fundamental change in effect in the boundary conditions imposed on the strongly non-linear plasma system might be expected intuitively to have the potential to affect its solutions accordingly. Nevertheless, a major difference in energy confinement behaviour first observed between all-C and ILW H-modes[85-87] had not been anticipated. Similarly, completely different responses to nitrogen seeding in ILW compared to all-C pulses were found, as outlined in a separate report[88]. Here we extend the analysis of these results, concentrating particularly on recycling, ELM, divertor-detachment, stationarity and especially power-balance aspects.

The experimental arrangement adopted and some elements of the treatment pursued throughout are first presented, before giving, for completeness of context, a short summary of contrasting confinement and underlying pedestal features, adding details not hitherto described. This is followed by comparison of inter-ELM recycling in the two JET designs, highlighting the pronounced departure in temporal dynamics and hence in majority boundary conditions which does indeed emerge. Performance of Type I H-mode is inevitably governed by ELMs themselves, however, and their variation under fuelling/seeding scans is next examined, especially with regard to their correlation with normalised confinement. A new region of falling ELM frequency for rising fuelling in the ILW is identified, while its confinement is shown actually not to depend strongly on their repetition rate and even to be positively correlated with mounting core radiation. As explained, the principal aim of impurity seeding is to moderate divertor heat loads, which are then investigated in the context of a thorough power-balance analysis both between and for ELMs. Reduction of both components in proportion to ILW N-seeding effects is revealed, while as a corollary, energy-amplitude scaling of the latter transients is also checked. This leads naturally into the approach to divertor detachment, for ultimate dispersal of plasma exhaust, the beneficial influence of N on which is considered for the ILW pulses. An equally important criterion for successful H-mode states is preservation of adequate plasma purity, so W sources from the ILW outboard target are next inferred from visible spectroscopy and their impact on confined-plasma behaviour estimated from evolution of core total radiation and soft X-ray (SXR) emission. The non-stationary properties observed remain the key limitation of ILW operation so far. Finally, the several contrasting features of all-C and ILW Type I H-modes are summarised and overall implications for further development and exploitation of integrated impurity-seeded "baseline" scenarios in preparation for ITER are briefly discussed.

## 2. Experimental set-up and time-averaging

To try to exploit the domain of best H-mode confinement at high density found in the all-C version of JET[77-79], incremental changes of deuterium fuelling and nitrogen seeding were executed in respective series of high-triangularity (average $\delta \approx 0.4$, elongation $\kappa \approx 1.7$, inverse aspect ratio $\varepsilon \approx 0.3$), single-null divertor plasmas at 2.65 T, 2.5 MA, $q_{95} \approx 3.5$ in each of its wall designs. Closeness of the equilibrium reproduction in both environments is confirmed by the EFIT[89-91] reconstructions during current and heating flat-tops of sample shots in Fig. 1, where the poloidal cross-section of the ILW is also depicted. In this variant, the strike-points in particular land on a tungsten-coated tile at the inboard side and on a bulk-tungsten target composed of four toroidal stacks of insulated lamellae[92-95] at the ouboard side. The inlet position of nitrogen directly into the outboard scrape-off-layer (SOL) plasma from the floor of the divertor is superimposed. Additional heating was primarily from neutral-beam injection (NBI), supplemented just in the all-C pulses with 1‑2 MW of ion-cyclotron-resonance-frequency (ICRF) waves to help regularise sawtooth oscillations[96]. Technical difficulties prevented ICRF power being included so far in ILW studies, possibly with some repercussions on their impurity and sawtooth behaviour, as considered further below.

Comparable matrix-like coverage of the two gas inputs was thus obtained in each environment, as plotted in Fig. 1 (all-C data are always shown as open symbols, ILW data as filled ones). Here and throughout, respective rates $\Phi_D$, $\Phi_N$ in each shot are characterised by the average number of electrons added per second during flat-tops, assuming full ionisation, taken to measure the influence on steadiest properties. Note this is preferred to a nitrogen spectroscopic radiance[97] owing to the variation in edge plasma temperatures, and hence ionisation to emissivity ($S/(XB)$) coefficients[73], involved. Wide ranges $0.75 \leq \Phi_D \leq 3.3$, $0 \leq \Phi_N \leq 4.7$ ($10^{22}$ el/s) were spanned, in order to pursue conditions from near natural H-mode density[98] to divertor detachment. Also displayed in Fig. 1 is the colour coding of points used throughout to indicate low (0.5‑1.0), medium (1.0‑2.0) and high (2.2‑3.3) D fuelling levels ($10^{22}$ el/s). All pulses were well above the L‑H transition threshold[87,99], placing them clearly in the Type I ELMs regime[100] in the absence of seeding or strong fuelling, with most lying between 15‑17 MW.

Attention is focused in following studies primarily on signals averaged over time during segments of flat-top phases. To indicate the degree of steadiness, "error bars" are then defined to express variability ($\sigma_\pm$) about these mean values, rather than actual measurement uncertainties. An important point is that such temporal variations are generally not normally distributed, so wherever time traces are averaged throughout this paper their variance is quantified separately above and below the mean :-

$$\bar{f} = \left[\sum_{k=1}^{N} \int_{t_{1k}}^{t_{2k}} f(t')\, dt'\right] \bigg/ \left[\sum_{k=1}^{N} \{t_{2k} - t_{1k}\}\right] \, ,$$

$$\sigma_\pm^2 = \left[\sum_{k=1}^{N} \int_{t_{1k}}^{t_{2k}} \Theta(\pm\{f(t') - \bar{f}\})\{f(t') - \bar{f}\}^2\, dt'\right] \bigg/ \left[\sum_{k=1}^{N} \{t_{2k} - t_{1k}\}\right] \, ; \quad \Theta(x) \equiv \begin{cases} 1, & x \geq 0 \\ 0, & x < 0 \end{cases} ,$$

(1)

where $\sigma_+^2 + \sigma_-^2 = \sigma^2$. This distinction is necessary particularly to discriminate where averages are significantly distinguished from zero and for clarity is therefore indicated by the label "non-Gaussian errors" (NGE) in every figure where it applies. The number $N$ of time intervals of signal $f$ considered may be 1 where a complete flat-top window is selected or >1 when only portions explicitly between ELMs are included. In this latter situation, recovery phases typically seen in each signal following ELM crashes[42,43] are largely avoided by choosing intervals 50‑90% of the way between consecutive transient peaks $t_{1k} = \{t_k^{ELM} + t_{k+1}^{ELM}\}/2$; $t_{2k} = \{t_k^{ELM} + 9t_{k+1}^{ELM}\}/10$ (see Fig. 7(a)). Consideration of such inter-ELM averages is particularly suggested since impurity seeding is apt mainly to affect power exhaust during these phases [31,35,38,101], consequently emphasizing their reactions and balances first.

## 3. Summary of all-C versus ILW energy-confinement, density and radiation effects

To provide the essential background for subsequent sections, key differences observed between high-triangularity H-mode plasmas in the two JET wall designs are briefly reviewed, further extending a first account given before [88].

### 3.1 Unseeded confinement

The subset of unseeded plasmas included (along the *x*-axis) in Fig. 1 captures the contrast in underlying behaviour of Type I H-modes in the all-C and ILW JET contexts. These pure fuelling cases are picked out in Fig. 2, using only in these two figures different symbols to convey total input power levels. Performance during pulse flat-tops is assessed in terms of energy confinement from EFIT [89-91] reconstruction, normalised to the ITERH98(y,2) scaling law [102], as a function of line-averaged density inferred from high-resolution Thomson scattering (HRTS) [103,104] profiles, in turn normalised to the Greenwald limiting value [66]. Quantities are time-averaged over a 1 - 2 s window during the steadiest phase of each shot, according to (1) ($N = 1$); temporal variances in Fig. 2 are small, signifying relatively steady signals.

The open symbols representing all-C data in Fig. 2 demonstrate the capacity, referred to above, to reach high densities up to or even slightly beyond the Greenwald limit by deuterium puffing, without any significant loss of normalised confinement [42-44,77-79]. As mentiond later, this was actually associated with a special behaviour of the ELMs [78,79]. Against this, for matching fuelling rate in the ILW (filled symbols), density stays ~10% lower, while in spite also of matching configuration and input power, normalised confinement is ≈ 30% lower and decreasing with density. Altered recycling in the ILW, noted above, might be thought capable of leading to different neutral-particle abundances around its plasmas, a property known to be able to affect confinement quality [24,105], but contours of accompanying torus gas pressure measured by a Penning gauge, superimposed ‡ in Fig. 2, suggest even this quantity was similar between the two datasets. Strictly this does not rule out the possibility still of different neutral-particle densities and mean energies, hence different edge source rates, between the all-C and ILW plasmas. Nevertheless the figure exemplifies the key departure in H-mode confinement between pulses with equivalent operational parameters in the two JET versions [85-88], which as seen below really stems from a reduction in (electron) pedestal temperature in the ILW [87,88], yet to be explained.

Over the range in input power (different symbols) spanned in Fig. 2, the smaller attendant variation in global properties implies the margin above L - H threshold [87,99] remains sufficent here for heating to be less influential than other factors, such as fuelling. Power dependence is therefore not considered further in this paper and attention is concentrated hereafter on the largest group of shots between 15 - 17 MW.

---

‡ That is, contours are interpolated between the torus-pressure values for each all-C and ILW shot at its position in the normalised confinement versus density plane.

## 3.2 Density and radiation with N seeding

A crucial feature of previous all-C plasmas was that while fuelling raised their density, N seeding tended to lower it again [42,43]. As just observed, in the ILW density was initially lower for equal fuelling of unseeded pulses, but on the contrary it then tended to rise as N was injected. This is evident in Fig. 3 (left) where time-averaged Greenwald fraction during flat-tops is plotted versus seeding rate. Note that in this and all subsequent scatter plots, different symbols are now used to denote intervals of N input; in particular, circles are always unseeded. Similar high density above the Greenwald limit could therefore be recovered in the ILW, but only with strong seeding.

These opposite responses of the density to N had important consequences for total radiated power, determined from bolometry covering the whole plasma including X-point and divertor regions, also shown in Fig. 3 (right). As explained above, this is averaged over intervals 50 - 90% of the way between neighbouring ELM peaks during a 1 - 2 s flat-top window, which are combined as in equation (1). Resulting inter-ELM total radiation fraction $f_{rad}^{i-E}$ is then displayed against plasma purity measured by line-averaged effective ionic charge $Z_{eff}$, derived from visible bremsstrahlung emission and time-averaged over the whole flat-top window including ELMs, since these tend to dominate in intrinsic impurity sources, as seen later. The decrease in density with seeding in all-C cases could actually lead initially to a fall in $f_{rad}^{i-E}$ for mounting N input [42,43], before it rose again at highest seeding, as best seen for higher fuelling (open blue symbols) in Fig. 3. In the ILW, an expected improvement in purity, owing to an at least ten-fold reduction in intrinsic carbon content [106,107], together with their lower density, yielded lower $f_{rad}^{i-E}$ in unseeded plasmas. Thereafter, increasing N injection was correlated with a uniform increase in $f_{rad}^{i-E}$, though the maximum value of $\approx 0.6$ reached still did not exceed that achieved at highest seeding in the all-C scans. A similar number for maximum inter-ELM radiation fraction had in fact been noted before in separate JET N-seeding studies [101], but whether it represents a genuine physical ceiling remains unclear. Classically, a 1-D approximation of edge power balance has been shown to lead to a maximum radiative loss for a given impurity species which depends only on upstream plasma density, its concentration and its average emissivity as a function of electron temperature convolved with the distribution of temperature and charge states [108]. However recent calculations [109] for JET-like conditions and accounting for enhancement by departures from coronal equilibrium [110] suggested the limiting radiation for N should still be much higher than the fraction of $\approx 0.6$ above. Whether this relates to a confinement-regime boundary [101] rather than practical saturation therefore remains to be tested. It may be noted finally, though, that strongest N seeding in the ILW only brought $Z_{eff}$ up to the level in the purest all-C pulses at higher fuelling.

## 3.3 Electron pedestal properties

It has been well established on JET and other tokamaks that global energy confinement in Type I H-mode is closely correlated with height of the edge pedestal [38,76,80,111-114], formed by its characteristic local reduction in transport [115-118]. In both JET designs, the electron component has been examined using data from the horizontal HRTS system [103,104], which captures effectively instantaneous 63-point profiles offering $\approx 1$ cm edge resolution at a rate of 20 Hz. Technical problems impeded detection of reliable charge-exchange recombination spectroscopic (CXRS) signals during ILW operations, so accompanying analysis of ion pedestals has not yet been available. During ELM cycles, pedestal profiles obviously vary continuously in time as they rebuild themselves at first rapidly after each crash and then more gradually until just before the next [119], meaning heights and shapes representative of properties near the stability limit [120-125] are realised in the later part of each inter-ELM period. In order to sample such profiles while acquiring adequate statistics, HRTS measurements have been chosen where they fall 65 - 95% of the way between consecutive ELM peaks (see §5) during a 1 - 2 s flat-top window and then averaged at each spatial position. Since this involves sets of smaller, discrete populations rather than segments of finer time-traces as in (1), an "error bar" conveying associated variability is estimated simply by symmetric standard deviation. Example temperature and density functions mapped to mid-plane major-radial co-ordinate $R$ are illustrated in Fig. 4(a) for three pulses from the ILW scans. To characterise the conspicuous pedestals, it is convenient to introduce a standard parameterisation quantifying their features in a consistent way, and the commonly-adopted form of a modified hyperbolic tangent [126-129] has been applied :-

$$F(R) \;=\; b \;+\; \frac{h}{2}\left[\tanh\!\left(\frac{R_c - R}{d}\right) \;+\; 1\right] \;+\; m\left[R_c - d - R\right]\Theta(R_c - d - R) \quad, \tag{2}$$

where $\Theta$ is the Heaviside unit step defined in (1), $R_c$ is the central position for width $2d$, inner slope $-m$ for $R \leq R_c - d$ and total height $b + h$. Electron density, temperature and pressure have each been interpolated with (2) for normalised minor radius $\rho \geq 0.85$ using a non-linear least-squares fitting procedure with Gaussian errors, giving e.g. the bold lines superimposed in Fig. 4(a). Note this treatment neglects convolution with the HRTS instrument function [129], but the effect is expected to be smaller on pedestal height, considered here, than on its width.

An interesting observation is that Fig. 4(a) immediately implies persistent flatness of the density and almost fixed gradient of the temperature across the main conduction zone between the sawtooth region and pedestal top. As N is added for only a modest change in fuelling, the density pedestal first increases substantially in height, then more strikingly the temperature pedestal rises for a somewhat lower density again. This remarkable response to N seeding is reminiscent of a similar finding for improved H-mode, or "hybrid", plasmas at high renormalised pressure [130] ($\beta_N$) in the all-tungsten ASDEX Upgrade machine [25-27] but in these experiments is seen on JET for the first time [88]. A comparison of full ILW against all-C scans is presented in terms of interpolated temperature versus density pedestal heights in Fig. 4(b). In all-C plasmas, their ability to reach high density and confinement together [42-44,77-79] is again revealed by highest pedestal pressure being achieved at higher fuelling, while N seeding tends to reduce it again mainly by the loss of density already described. In contrast, ILW plasmas can be seen to span a similar range in density, but to start in unseeded pulses at significantly lower temperature; in fact, Type I regime persists here at a value where Type III ELMs would have been expected in the all-C device [131]. The capacity of N seeding in the ILW initially to raise density, typified in Fig. 4(a), then pushes its pedestals to similar high pressure as in all-C instances, thereafter sustaining this condition as next the temperature rises for commensurately falling density. In the best cases, N-seeded all-C and ILW pedestals are thus roughly overlapping. Finally strongest N input leads to a loss of pedestal pressure again (blue cross), suggesting a most favourable state has been passed. In addition, it is apparent that highest ILW pedestal temperatures (though not pressure) are realised at lowest fuelling level, but these plasmas are conversely among the least steady, as explained later in §10.

It therefore appears that pedestal stability can be improved, in the sense that higher top values are reached before an ELM is provoked, by the presence of impurities, either through their higher charge, or mass, or both. In the case of all-C plasmas, with higher intrinsic $Z_{\text{eff}}$, it is conceivable this role has been filled by carbon, so that as first proposed elsewhere [88], it played a hitherto unrecognised part in their high pedestals and robust confinement. Detailed modelling to explore this question is continuing. The correlation of normalised energy confinement with electron pedestal pressure is plotted in Fig. 5, where the tendencies become clear for each to fall with seeding in all-C pulses, but to rise with it in ILW ones, until similar values are realised in both environments. Note, however, that ILW plasmas never recover completely the best performance of unseeded all-C states. Also superimposed are best proportionality lines separately through the sets of all-C and ILW points, showing very nearly the same direct relationship between pedestal height and confinement is preserved in both contexts. In other words, confinement is generally lower in ILW H-modes, as in Fig. 2, because the pedestal is lower, in turn because the pedestal temperature is lower, the cause of which, as remarked, remains the subject of on-going study. Furthermore, the new ILW aptitude for normalised confinement to improve with N seeding, not observed in earlier JET designs [35,38], arises from the gain elicited in pedestal height. Such a conclusion is further reinforced in Fig. 5 by the normalised inverse scale-length of electron pressure $R\, \partial(\ln p_e^{\text{i-E}})/\partial R$, approximated from a least-squares linear fit to the same inter-ELM average profiles over an interval in normalised minor radius $0.3 \le \rho \le 0.8$. This quantity even falls slightly as performance rises, particularly in the ILW, confirming confinement is not generally improved by any change in core peaking, but is indeed governed by the pedestal. Lastly, electron collisionality [132] evaluated at the inter-ELM pedestal top values is also included in Fig. 5 and interestingly displays rather little variation as the pressure height alters, staying close to the marginal level ($\nu_{*e}^{\text{ped i-E}} \sim 1$) throughout. Implications for ELM scaling are returned to below (§7).

## 4. Divertor recycling behaviour

Plasma interactions with boundary surfaces are apt to depend stongly upon the material involved, with carbon tending to be more adsorbing for both incident particles and energy, while metals, especially tungsten, are generally more reflective [81-84]. Replacement of the all-C lining by the ILW in JET was therefore likely to change its recycling behaviour profoundly, particularly in the divertor where plasma effluxes are concentrated. Typical features for high-triangularity Type I H-mode in the all-C environment are illustrated in Fig. 6, selecting an unseeded pulse at higher fuelling (#76678). Time-traces of plasma stored energy from fast magnetic reconstruction and total radiated power from bolometry disclose characteristic fast transients at each ELM and smooth recovery in between. Divertor recycling is depicted by Balmer-α (656 nm) radiance along filtered photomultiplier lines-of-sight intercepting the inboard (ISP) and outboard strike-points (OSP) and similarly shows prompt rises at both sides at each ELM, which, within measurement uncertainties, are simultaneous with each other and the energy crashes / radiation spikes. In between, recycling becomes steady again, with post-ELM relaxation at the OSP indicating a prevailing pump-out time. Peak heat load at the outboard side from fast ($\approx 12$ kHz) infra-red thermography [133] also exhibits a very sharp spike at each ELM, again coincident with those in all other signals, within errors. Although described here for an unseeded pulse, these features in fact remained qualitatively unchanged in every all-C plasma, whether without or with N seeding.

A close counterpart unseeded case in the ILW (#83357) reveals a radically different type of ELM cycle. Radiation spikes at each fluctuation are now much smaller, consistent with lower intrinsic impurity content, and are accompanied by slower drops in stored energy. This longer time-scale of energy loss is particularly evident in thermographic detection here of peak heat load on the divertor outboard target, the broader rise and fall of which at each transient implies a slower plasma unload [87], although maximum power densities reached remain close to those in the all-C example. Recall that as mentioned previously, the ILW outboard target is composed of four thermally-isolated toroidal stacks of solid W segments [92-95], labelled A - D from the inboard to the outboard side (see top-left inset in Fig.6), the heat load plotted being for stack C where the OSP lands. Comparable power densities on the ILW and all-C targets despite contrasting time-scales are then partly attributable to steeper field-line incidence, and lower toroidal wetted fraction, for the former. Balmer-α radiances at the ISP and OSP are complemented by additional lines-of-sight viewing the divertor-entrance baffles at each side, both of which, together with the ISP, now actually show a drop in recycling at each ELM. Only at the divertor OSP is there still a rise in $D_\alpha$ brightness, but this occurs after a noticeable delay of ~ 5 - 20 ms, depending on the ELM size. Equally striking is the appearance too of spontaneous oscillations in recycling between ELMs, at a frequency of ~ 200 Hz and in phase between the ISP and both entrance baffles but exactly in anti-phase at the OSP, though at much lower amplitude (see top-centre inset in Fig. 6). They are also local to the divertor, with no significant impact on the main plasma, as witnessed by their absence from the radiation and stored energy signals. Divertor oscillations at much lower frequency had sometimes been seen before on JET but only in L-mode and with no localisation to its vicinity[75,134]. Such localisation and apparent in-out alternation in ILW recycling, moreover, are not obvious aspects of classical vacillation between bistable higher and lower edge density states, well known theoretically[135-137]. Roughly ten times higher $D_\alpha$ radiance in the ILW divertor than in the all-C one for otherwise similar global parameters is consistent with more reflective targets (i.e. a higher recycling regime) in the former and hints at the possibility that self-sustained oscillations might be damped in the latter by stronger adsorption at carbon surfaces. Further study is required, however, to elucidate the exact nature of these now common ILW recycling dynamics.

Unlike the all-C situation, the markedly altered ELM and divertor phenomena in ILW unseeded plasmas are not robust against changed inputs and are readily affected by N seeding. A pulse at lower fuelling but stronger N dosing (#82813) in Fig. 6 demonstrates that its admission increases the radiation amplitude of ELMs again (as well as inter-ELM emission, Fig. 3), while tending to restore the faster time-scale of energy collapse, resembling all-C instances[87]. Divertor oscillations are almost immediately suppressed too, which, accounting for reduced efflux power due to increased radiation (Fig. 3), could in fact be consistent with a power threshold predicted for bistability fluctuations[135,136]. Recycling at the ISP and both sides of the divertor entrance now exhibits sharp, coincident bursts akin to the all-C transients, whereas a simultaneous drop occurs at the OSP, thereby reversing the pattern of the unseeded state. Such inversion emerges particularly when N seeding exceeds the fuelling rate ($\Phi_N > \Phi_D$), indicating the strong effect extrinsic species can exert on majority processes. The exact mechanism of this interplay is being studied through 2-D edge modelling, though a strong reduction in shot-to-shot N legacy[138] (i.e. incomplete removal in between) compared to the all-C divertor already points to higher recycling of N itself in the ILW. Persistence of the in-out asymmetry at ELMs throughout the scans means, however, that the ILW divertor never reverts to the more balanced recycling of its all-C predecessor.

These swingeing changes in recycling hint at different detachment effects, considered further below, between the two JET environments but more fundamentally signify a pronounced alteration effectively in the boundary conditions imposed on their respective plasmas. For such a strongly non-linear system, this could potentially affect their behaviour accordingly and, together with reduced intrinsic impurity concentrations, intuitively seems a likely factor in contrasting ILW performance.

## 5. ELM properties and normalised confinement

It has already been seen that different pedestal responses in the ILW underlie its modified plasma confinement in fuelled and N-seeded pulses. A corollary and signature of changing pedestal stability is therefore the variations in ELM fluctuations[124,139,140] which punctuate steadier H-mode states. The nature of ELM occurrence tends to resemble a statistical process[78,141-144] and in any scheme seeking to identify them from time-series data generally a continuous spectrum of structural and temporal deviations will be encountered in a sufficiently wide sample. Any simple algorithm thus tends to become arbitrary at some level, so that a standard means suitably to characterise their form is still being sought[142-144]. Consequently for these experiments, only a basic level-crossing scheme has been applied to time-traces of total radiated power fraction from bolometry, examined over a 1‑2 s window during pulse flat-tops. An example of resulting identification of ELM peaks is shown for an unseeded ILW case at higher fuelling in Fig. 7(a), where every transient selected is labelled by a dot. The ambiguity intrinsic to such analyses is typified by the inset redrawing $f_{rad}$ around 15.25 s on a greatly expanded time-scale, from which it is clear that choosing a lower onset threshold could have detected two peaks rather than one, separated by a very much shorter interlude. As noted, ambiguities of a similar kind will generally persist for any choice of threshold or any equivalent series criterion, so ELM events derived cannot yet be regarded as completely objective. Nevertheless dominant trends of most prominent instabilities should be obtained.

**5.1 Frequency distributions**

Times of each ELM peak have first been used to define the set of inter-ELM periods $[t_{1k}, t_{2k}]$; $k = 1,...,N$ in each pulse over which its time-signals have been averaged according to (1), taking those portions 50 - 90% of the way between ELMs to avoid main post-crash recovery phases [42,43], as stated above. In addition, the reciprocal of the full interval between each pair of consecutive peaks defines a momentary frequency, superimposed in Fig. 7(a), from a sequence of which a sample distribution can hence be constructed. These have been accumulated with a fixed bin size of 5 Hz and normalised to give approximate frequency probability distributions, as illustrated for all-C and ILW unseeded and N-seeded plasmas at higher fuelling in Fig. 7(b). Each example reveals not only that ELMs can very rarely be characterised well by a single number for their frequency, but also that their considerable variability is seldom normally distributed. On the contrary, their frequency distributions tend to be wide and distinctly asymmetric. In both JET variants, their breadth also increases when N is added.

It is conventional to measure a distribution by its mean and this has been estimated using a discrete analogue of (1) to account still for asymmetry :-

$$\bar{\nu} = \sum_j r_j \nu_j \quad ; \quad r_j = \eta_j / \sum_j \eta_j = \eta_j / N \quad ,$$

$$\sigma_\pm^2 = \left[ \sum_j \Theta(\pm\{\nu_j - \bar{\nu}\}) r_j \{\nu_j - \bar{\nu}\}^2 \right] \Big/ \left[ 1 - \frac{1}{N} \right] \quad , \tag{3}$$

where $\eta_j$ is the number of occurrences of frequency $\nu_j$ in the number $N$ of inter-ELM periods. However, this moment is easily affected by outliers, whereas the mode, i.e. $\nu_j$ at which approximate probability $r_j$ is maximum, by picking out the dominant frequency becomes less susceptible to extrema or subjectivity in the identification scheme. Both moments have therefore been used to quantify changes in ELM features.

Reaction of mean and modal ELM frequencies to N seeding in each JET design is summarised in Fig. 7I. Unseeded all-C plasmas first reveal the special response which emerged at high triangularity, viz. frequency of their ELMs actually decreased with rising fuelling [78]. Such "retrograde" behaviour, opposite to that found for lower shaping [78,100], was associated with enhanced inter-ELM effluxes and thereby underlay their preservation of a higher pedestal and confinement at densities up to the Greenwald limit [79]. Once N was added, however, ELMs plainly tended to become more frequent again, accompanied by a gradual decline in confinement [44]. In complete contrast, unseeded ILW pulses show their ELM frequencies now tend to increase with fuelling even at high triangularity, so that as implicit in Fig. 2, the exceptional "retrograde" regime of all-C results has not yet been recovered. On the other hand, injecting N initially causes a fall in ELM frequency, before it increases at higher seeding rate, hence defining a minimum at a certain N level. Within the rather coarse scan conducted it is feasible this pivotal N rate depends somewhat upon the simultaneous D fuelling level, leading to a new region between $2 < \Phi_N < 3$ ($10^{22}$ el/s) where the frequency of ELMs appears to reduce for mounting fuelling $\Phi_D$ (filled squares). Whether this might be associated with strengthening inter-ELM losses as in the all-C environment remains unclear, but it has already been seen in Fig. 4(b) that electron pedestal pressure for these pulses does tend to reach higher values with more fuelling before becoming unstable.

## 5.2 Correlation of confinement with ELM frequency

The tendency of ELM frequency first to fall with N seeding in the ILW naturally prompts the question whether this might be correlated with recovery of energy confinement newly emerging in these studies. Confinement factor $H_{98y}$ over both series of scans is plotted versus mean and modal ELM frequencies in Fig. 8(a). Against either moment, all-C data (open symbols) clearly do exhibit an inverse correlation of normalised confinement with ELM frequency, particularly at higher fuelling, i.e. $H_{98y}$ drops as $\nu_{ELM}$ increases. Such inverse dependence is entirely in line with conventional Fishpool scaling [119], which essentially describes higher time-average pedestal, and so confinement, values due to longer periods spent nearer to upper limits set by stability as ELMs become farther apart in time. In other words, enhanced performance stems from a combination of greater stability, allowing higher pedestals to be achieved, and dwelling longer close to these bounds, owing to more rapid recovery of edge profiles immediately after an ELM crash [119], as fluctuations intervene less frequently. Some similar reciprocal correlation is also evident for ILW counterparts (filled symbols) at higher fuelling, but it becomes progressively weaker as fuelling is lowered. Eventually, highest ILW confinement, realised as expected at lowest fuelling, becomes effectively independent of ELM frequency. Comparison of Figs. 5 & 8(a) discloses relatively little change in ILW electron pedestal pressure at lowest fuelling for rather a wide variation in either mean or modal $\nu_{ELM}$, with highest $H_{98y}$ occurring for most frequent ELMs on either measure. Small differences in confinement for these plasmas are in fact broadly consistent with corresponding changes in core peaking $R\, \partial(\ln p_e^{i-E})/\partial R$, also included in Fig. 5. The liability of N seeding to yield higher pedestals and confinement in the ILW while departing from simple Fishpool scaling therefore suggests it may genuinely be improving edge stability, and so raising critical pressure before an ELM is precipitated, as inferred above.

## 5.3 Confinement improvements with increased core radiation

Preceding details have established that density, radiated power (Fig. 3) and normalised confinement (Fig. 5) all rise with N seeding in the ILW. An equally interesting question to the correlation with ELM frequency is therefore that between the latter two quantities. Radiation from the confined plasma has been estimated from tomographic reconstructions using multiple bolometric lines-of-sight and timed to sample only the later stages of intervals between ELMs. Emission from the volume inside the separatrix (here denoted the "core") has then been averaged within a 1 - 2 s flat-top window, retaining asymmetric variances above and below the mean, analogously to (1). Alteration of energy confinement with N seeding in the ILW is plotted as a function of resulting inter-ELM core radiation fraction in the lower pane of Fig. 8(b), confirming that both properties indeed increase together, particularly at higher fuelling when initial unseeded values are lower. Thus radiation from the core plasma does not detract from its confinement in the ILW environment. Such a surprising feature is especially notable since the usual definition of energy replacement time $\tau_E$, adopted here, does not discount core radiation $P_{rad}^{core}$, which can in effect act like a confinement loss term :-

$$H_{98y} \equiv \frac{\tau_E}{\tau_{98y}} = \frac{W_{th}}{P_{in} - \dot{W}_{th}} \frac{1}{\tau_{98y}} \quad , \tag{4}$$

where $W_{th}$ is thermal stored energy, $P_{in}$ total input power and $\tau_{98y}$ confinement time predicted by the ITERH98(y,2) scaling law [102]; hence if $P_{rad}^{core}$ increases, $W_{th}$ and so $H_{98y}$ will be disposed to decrease, even if $P_{in}$ and $\dot{W}_{th}$ do not change. The fact that this does not happen in Fig. 8(b) suggests consideration of a modified definition of confinement focusing specifically upon plasma transport losses $P_{\text{loss}}^{\text{core}} = P_{\text{in}} - \dot{W}_{\text{th}} - P_{\text{rad}}^{\text{core}}$ across the closed magnetic surfaces, viz. :-

$$\tau'_E = \frac{W_{\text{th}}}{P_{\text{in}} - \dot{W}_{\text{th}} - P_{\text{rad}}^{\text{core}}} \quad \Rightarrow \quad H'_{98y} = H_{98y} \frac{W_{\text{th}}}{W_{\text{th}} - \tau_E P_{\text{rad}}^{\text{core}}} \quad , \tag{5}$$

where for comparison purposes normalisation has been made to the same scaling law, even though ITERH98(y,2) itself is based upon data for $\tau_E$ subsuming core radiation[102]. Alternative $H'_{98y}$ estimates averaged in 1 - 2 s flat-top windows are included in the middle pane of Fig. 8(b) and serve to emphasize that as core radiation between ELMs increases with N seeding in ILW plasmas, their underlying confinement of thermal energy improves dramatically. At the same time, inter-ELM fractional loss power crossing the separatrix $P_{\text{loss}}^{\text{core i-E}} / P_{\text{in}}$, given in the top pane of Fig. 8(b), barely changes within the square root of temporal variances. Here mean rates of change of stored energy have been approximated as outlined in next §6, but the inference which follows is that as $f_{\text{rad}}^{\text{core i-E}}$ rises with N seeding, the attendant adjustment of ELM occurrence is such that on average there is a roughly compensating fall in $\dot{W}_{\text{th}}^{i-E} / P_{\text{in}}$. Together with almost constant gradient across the main diffusive interval in minor radius of example electron temperature profiles already noted in Fig. 4(a), near invariance of fractional loss power would therefore be consistent with fixed average (electron) energy transport between ELMs over the ILW scans. The mechanism of strongly improving confinement $H'_{98y}$ indicated in Fig. 8(b) would then still appear to be principally the gain in pedestal height described above, although with a steeper proportionality than that seen jointly with all-C pulses for $H_{98y}$ in Fig. 5.

## 6. Power balance

The chief aim of impurity seeding has been explained to be moderation of divertor heat loads in order ultimately to achieve sustainable exhaust conditions in large, high-power tokamaks[2,12,14,17,18]. In the all-C version of JET, divertor-target power densities were derived from fast infra-red thermography[133], measuring high-resolution ($\approx 2$ mm) poloidal profiles typically at a framing rate of $\approx 12$ kHz[42,43]. For technical reasons, this diagnostic was only available for 2.5 MA discharges with N seeding in the ILW in some companion low-triangularity cases ($P_{\text{in}} = 16$ - 21 MW) and just three at high triangularity ($P_{\text{in}} = 16$ MW). Langmuir probes embedded within the surface of its bulk W outboard target were operational during most plasmas, however, and have been exploited instead to estimate incident heat fluxes. The foregoing small subset of instances with data from both instruments first allows a cross-check of their consistency, which is constructed in Fig. 9(a), incorporating just in this single figure low- as well as high-triangularity information. Note, though, that similar ranges in D fuelling and N seeding rates to those of the main high-triangularity scans (Fig. 1(a)) were thereby spanned. Respective IR-camera and probe integrals of power landing on the whole toroidal area of the divertor outboard target have again each been averaged over 50 - 90% segments of inter-ELM periods within 1.5 s flat-top windows, according to (1), accounting for the toroidal wetted fraction of $\varpi_t \approx 0.5$ applying to the former. For the latter, this factor is accommodated by taking the product of ion flux density mapped to a horizontal plane $\Gamma_{ih}^{\text{tgt}}$ with target-facing electron temperature $T_e^{\text{tgt}}$ at each probe, then integrating toroidally over the set covering all four stacks of the target :-

$$P_{\text{LP}}^{i-E} / \gamma = 2\pi \int R^{\text{tgt}} \Gamma_{ih}^{\text{tgt}} T_e^{\text{tgt}} ds \quad , \tag{6}$$

where $s$ is a rectilinear co-ordinate along the inclined tile surface and major radius $R^{\text{tgt}} = 0.968s + 1.566$ (m) in the ILW. Note the expression tacitly assumes probe currents consist entirely of deuterons, although in strongly seeded shots especially some contribution is likely also from extrinsic and intrinsic impurity ions. The right-hand side of equation (6) has dimensions of power but with the sheath total energy transmission factor[145-147] $\gamma$ omitted. Furthermore, Langmuir probe measurements exclude extra contributions to target loading from absorbed radiation and impinging neutral-particle fluxes, both of which would be expected to vary substantially over the fuelling / seeding scans and conversely to be included by IR thermography. Such effects may partly be responsible for substantial scatter visible in Fig. 9(a), but nevertheless it can be deduced the two diagnostics become broadly consistent assuming $\gamma = 8$. This commonly-used value[146,147] has therefore also been invoked hereafter for ILW inter-ELM target powers from probes. An illustration of the close agreement which can be obtained with IR profiles of power density deposited on the outboard target between ELMs $q_{\text{tgt}}^{\text{i-E}}$ (MW / m$^2$) is presented for one of the high-triangularity pulses in Fig. 9(b), where at each probe $q_{\text{tgt}}^{\text{i-E}} = 8\Gamma_{i\,h}^{\text{tgt}} T_e^{\text{tgt}} / \varpi_t$. The narrow gap between the two outermost toroidal stacks (C and D, cf. left-hand inset in Fig. 6) is noticed here as a spike in the IR signal at its illuminated edge. In addition, it is apparent the spacing of embedded probes yields resolution of $\approx 1.3$ cm along $s$ at best, constraining the accuracy of heat-load estimates examined below.

## 6.1 Main components of global power balance

For the pulse in Fig. 9(b), every probe within the ILW outboard target was operational, offering the most precise definition of incident plasma loads achievable with this system. In several shots, however, one or more probes failed to register data at locations which left significant uncertainty whether the peak power density had been captured, i.e. whether a profile had been reasonably well detected. Integral powers (6) for such cases can thus only be regarded as lower bounds and they have been separated out in following analyses by use of faint symbols on plots; remaining pulses with best probe data, adequately describing profiles and integral loads, are depicted by plain symbols.

As with all other temporal characteristics, the momentary power balance in Type I H-modes tends to be controlled by ELM occurrence. It has already been commented previously that the pedestal and hence global confinement recover continuously after each ELM crash until the next instability intervenes[119], meaning there is generally a positive change of thermal energy stored by the plasma $\dot{W}_{\text{th}}^{\text{i-E}}$. To estimate this term, fast (2.5 kHz) magnetic reconstruction of stored energy has been linearly interpolated by connecting instants 50% and 90% of the way between each pair of consecutive ELM peaks and then averaging the resulting derivative within 1 - 2 s windows during pulse flat-tops. Since this involves smaller sets of samples, as with HRTS profiles in §3.3, variability has again been approximated just by symmetric standard deviation. This effective reheating term is juxtaposed in Fig. 10 (left) with inter-ELM estimates of total radiation from bolometry and outboard target load derived from Langmuir probes, as outlined above, for the subset of pulses with best probe aggregate profiles, normalising each quantity to total input power. The surprising importance of $\dot{W}_{\text{th}}^{\text{i-E}} / P_{\text{in}}$ in unseeded plasmas is immediately evident, while even in N-seeded cases where ELM frequency, as seen, can decrease, it mostly remains a larger component than inter-ELM outboard-target heat load. The net balance represented by $(P_{\text{rad}}^{\text{i-E}} + \dot{W}_{\text{th}}^{\text{i-E}} + P_{\text{out.tgt}}^{\text{i-E}}) / P_{\text{in}}$ is also shown as a function of N input in Fig. 10 (right), including instances both with best probe data and those with some signals around the profile peak missing. Note that a matching estimate of total radiation was not obtained in the most strongly seeded shot (filled cross in previous scatter plots, approximated from a different signal in Fig. 2) and so for consistency it has been excluded from power-balance analyses. The foil bolometers in JET are likely to register also their local part of atomic fluxes, principally from charge-exchange events, but other components absent from overall power accounting in Fig. 10 include neutral-beam shine-through and fluxes directed onto other material surfaces, particularly the divertor inboard target. Allowing for these omissions, the main power-balance fractions between ELMs compiled in Fig. 10 seem reasonable across the entire ILW fuelling / seeding scans, even for those pulses with less complete probe datasets.

**6.2 Inter-ELM heat load and ELM losses**

It is already clear in Fig. 10 (left) that as inter-ELM radiation is increased with N seeding, accompanying power efflux onto the ILW outboard target can be markedly reduced. The variation is shown more systematically as a function of N input in Fig. 11 (top left), where the three corresponding measurements of target heat load from fast IR thermography are also superimposed, helping to corroborate the lower-resolution probe integrals. A definite decline in $P_{\text{out.tgt}}^{\text{i-E}} / P_{\text{in}}$ emerges for stronger N seeding ($\Phi_N > 2 \times 10^{22}$ el/s), becoming more pronounced at lower fuelling, for which it will be recalled ELM frequency is higher (cf. Fig. 7I). This drop is actually explained in the adjacent graph (top centre), where the same normalised target load is replotted against net plasma loss power fraction $f_{\text{loss}}^{\text{i-E}} = (P_{\text{in}} - P_{\text{rad}}^{\text{i-E}} - \dot{W}_{\text{th}}^{\text{i-E}}) / P_{\text{in}}$ (note this differs from the denominator in (5) since $f_{\text{loss}}^{\text{i-E}}$ discounts all radiation between ELMs, including that from the SOL and divertor). It thus becomes apparent, particularly from the subset of pulses with more reliable probe profiles, that $P_{\text{out.tgt}}^{\text{i-E}} / P_{\text{in}}$ falls in proportion to the reduction in $f_{\text{loss}}^{\text{i-E}}$, except possibly for the highest seeding discharge at lowest fuelling, a point returned to later. In other words, $P_{\text{out.tgt}}^{\text{i-E}} / P_{\text{in}}$ does generally decrease in proportion to the extra radiative cooling effect of N seeding, combined with its impact on ELM frequency and consequent plasma reheating rate $\dot{W}_{\text{th}}^{\text{i-E}} / P_{\text{in}}$ between crashes. Complexity of the latter response to fuelling and seeding is principally responsible for the non-uniform behaviour of $P_{\text{out.tgt}}^{\text{i-E}} / P_{\text{in}}$ against N input in the top-left plot.

An equivalent correlation in counterpart all-C scans is tested in the top-right pane of Fig. 11. Here target loading is determined from fast IR measurements, so will include also radiative and atomic heating of the plate not captured by the ILW probe estimates, although outboard-target and net loss power fractions do still lie in very similar ranges in both environments. Two all-C pulses at highest N seeding and lower fuelling stand apart from the remainder, but these had very high ELM frequencies (hundreds of Hz) and may have entered Type III regime [131]. Except for these two outliers, all-C variations in $f_{\text{loss}}^{\text{i-E}}$ and $P_{\text{out.tgt}}^{\text{i-E}} / P_{\text{in}}$ are less than those in the ILW, consistent with the smaller change of $f_{\text{rad}}^{\text{i-E}}$ in Fig. 2. Nevertheless, within their narrower intervals and comparable temporal variances, $P_{\text{out.tgt}}^{\text{i-E}} / P_{\text{in}}$ again decreases in proportion to $f_{\text{loss}}^{\text{i-E}}$, especially at lower fuelling. A similar conclusion had been reached from simpler energy-balance analyses before [43].

The obverse change to inter-ELM power exhaust is that for the ELM bursts themselves. Complementary to the response of ELM frequency to N seeding, summarised in Fig. 7I, is that of ELM average energy amplitude $\Delta W_{\text{ELM}}$, depicted in the lower-left pane of Fig. 11. Here only a leading-order approximation of $\Delta W_{\text{ELM}}$ has been inferred from the drop in plasma stored energy at each transient using fast magnetic reconstruction, averaged within the same flat-top windows with symmetric standard deviations, rather than invoking more thorough but demanding integration of high-time-resolution profiles [87,148]. Separation of points into those with best and partially-missing probe data has also been retained to facilitate comparison with other plots, although it obviously does not affect the ordinate concerned. Even this rough estimate of ELM size, however, demonstrates a clear tendency for $\Delta W_{\text{ELM}}$ to increase reciprocally as fluctuations become less frequent under N seeding, before diminishing again as ELM frequency is restored at higher N rates. Largest amplitudes reach $\approx 700$ kJ, roughly twice the maximum size of ELMs realised in the partner all-C scans (not shown), for unseeded plasmas. Note absolute, rather than normalised, values are examined since these relate directly to W erosion in the ILW divertor, considered later.

Dividing each ELM energy drop by its immediately preceding inter-ELM period implies an associated power, which has again been averaged over the chosen flat-top window in each pulse. This effective mean power exhaust by ELMs, normalised to input power, is finally plotted versus net inter-ELM loss power fraction $f_{loss}^{i-E}$ for the ILW in the bottom-centre, and all-C equivalents in the bottom-right, panes of Fig. 11. A simple connection between ELM losses and intervening exhaust would not necessarily be expected a priori and such a view is borne out in the all-C cases, where $P_{ELM}/P_{in}$ falls only weakly with rising N input, reminiscent of an earlier conclusion[37,38] that Type I ELMs tend not to be substantially moderated by extrinsic radiation. Discounting the two pulses encroaching on Type III ELMs, for which $P_{ELM}/P_{in}$ is effectively undetermined, there is otherwise no strong correlation with $f_{loss}^{i-E}$, even an inverse trend being suggested of rising $P_{ELM}/P_{in}$ as $f_{loss}^{i-E}$ falls. In marked contrast, though, ILW plasmas exhibit a pronounced drop in $P_{ELM}/P_{in}$ as N is added, at all fuelling levels, which furthermore is generally in close proportion to, or even slightly steeper than, the decrease of $f_{loss}^{i-E}$. In other words, somewhat unexpectedly, ELM losses are more strongly affected by N seeding in the ILW and again reduce approximately in line with its cooling and ELM-cycle effects, adding an extra large benefit to its use. This further wide departure between behaviour in the all-C and ILW environments provides additional evidence of the altered pedestal properties and dynamics prevailing in the latter. Carefully comparing upper and lower plots against $f_{loss}^{i-E}$ in Fig. 11 reveals that for either choice of wall materials, however, average power fraction ejected by ELMs almost always exceeds that deposited on the outboard target in between, with $P_{ELM}/P_{in}$ only approaching $P_{out.tgt}^{i-E}/P_{in}$ at highest N seeding in the all-C scans, and only becoming less than it for lower N seeding at medium and higher D fuelling in the ILW, when ELM frequency touches a minimum (cf. Fig. 7I). Again this change in the balance of power efflux demonstrates the capacity of extrinsic seeding fundamentally to influence the character of ILW H-modes. More generally, the persistence of a dominant ELM component in time-average exhaust re-emphasizes the probable need ultimately to combine extrinsic seeding with some form of active mitigation[149-162] of Type I fluctuations.

## 7. ELM size scaling

An estimate of electron collisionality[132] at the pedestal top, during the later stages of ELM cycles, has already been derived from HRTS measurements in Fig. 5. A broad set of unseeded-plasma data from several tokamaks[141] had previously indicated this quantity tends to be a leading factor in determining the energy amplitudes of Type I ELMs when expressed as a fraction of that stored in the pedestal just prior to each crash, viz. a definite inverse correlation was displayed over 2‑3 orders of magnitude in $\nu_{*e}^{ped\,i-E}$. The preceding values for $\Delta W_{ELM}$ have therefore been used to check such a relation over the fuelling / N-seeding scans, normalising them to approximate pedestal energy $W_{ped}^{i-E} \approx 3 p_e^{ped\,i-E} V_p$, where $V_p$ is the confined plasma volume from EFIT[89-91] reconstruction and assuming $p_i^{ped\,i-E} \approx p_e^{ped\,i-E}$ in the absence of ion temperature / density information. Resulting fractional losses are shown as functions of $\nu_{*e}^{ped\,i-E}$ for each JET environment in Fig. 12. Although the derivation of $\Delta W_{ELM}$ is rather crude here, as explained above, normalised sizes still span a similar range ($\sim 5\text{-}30\%$) to that in the original survey[141]. Moreover, they are actually slightly higher in unseeded ILW than matching all-C plasmas, despite much smaller ELM radiation amplitudes for the former as exemplified in Fig. 6, owing to their significantly lower pedestals. In both JET variants, N seeding does then lead to a genuine reduction of ELM energy drops $\Delta W_{ELM}$ (cf. Fig. 11 bottom left for ILW) such that, combined with governing pedestal changes, $\Delta W_{ELM}/W_{ped}^{i-E}$ decreases by up to a factor of 3 in all-C, and by up to a factor of 5 in ILW, sequences at highest fuelling. In the ILW, intermediate seeding and fuelling cases again stand out as they realise a minimum in ELM frequency and so a maximum in energy amplitude.

The rather wide interval in normalised ELM size covered is not quite matched by that in pedestal collisionality. In fact, $\nu_{*e}^{ped\,i-E}$ varies only by a factor of $\approx 2$-$3$, similar to that in electron pressure pedestal height for both wall environments combined (cf. Fig. 5), and as remarked in §3.3 above stays close to the marginal level $\nu_{*e}^{ped\,i-E} \sim 1$ throughout. Whether this intriguing feature is coincidental, or indicative of other undisclosed effects within the edge plasma, has yet to be elucidated. Notwithstanding the narrow range in collisionality spanned, however, it is clear in Fig. 12 that an inverse correlation of $\Delta W_{ELM}/W_{ped}^{i-E}$ with $\nu_{*e}^{ped\,i-E}$, according with earlier scaling [141], is not in evidence for either all-C or ILW datasets. This departure seems particularly noteworthy for the all-C pulses, since a major part of the original study itself consisted of high-triangularity H-modes at 2.5 MA in a version of JET with all-C walls [141], albeit with a different divertor geometry and excluding extrinsic seeding. Moreover, it also contradicts separate seeding results, using argon, in the same intermediate divertor design of the JET all-C machine [36]. Further investigation is required to determine whether a new domain of pedestal behaviour might consequently be signified. In the ILW, though, it should be recalled from above (§3.3) that a new regime of Type I ELMs is already suggested by their appearance in unseeded plasmas at pedestal temperatures well below the boundary for onset of Type III properties in all-C experiments [100,131].

## 8. Outboard detachment

At sufficiently low divertor-plasma temperatures, volume processes can dominate over transport within the SOL region and quench charged-particle fluxes to material targets, instead spreading power, momentum and particles over larger enclosing surfaces, to yield an optimum exhaust regime of detached operation [75]. Access to such conditions is consequently a high priority also for impurity-seeded H-modes and has expressly been sought in fuelling/seeding scans in both first-wall versions of JET. Here the approach to detachment at the outboard target in ILW pulses is particularly considered, in order to help illuminate some of the effects of an extrinsic species on its development.

It is useful first to recall how volume processes can affect steady balances along a classical SOL plasma during the preliminary stages while it remains attached. Neglecting cross-field terms, viscous stress and changes in total induction $B$, steady 1-D currentless force balance may be integrated between upstream (mid) and target (tgt) ends of the SOL flux-tube to give [74,163] :-

$$\frac{p^{tgt}}{p^{mid}} \approx \varphi_p \frac{1+(M_\parallel^{mid})^2}{1+(M_\parallel^{tgt})^2} \approx \varphi_p \frac{1}{1+(M_\parallel^{tgt})^2} \quad , \tag{7}$$

where $p \equiv p_e + p_i$ is total scalar pressure, $M_\parallel$ is the Mach number of parallel flow and typically $(M_\parallel^{mid})^2 \ll 1$, while $0 \le \varphi_p \le 1$ is a friction factor conveying the impact of volume momentum sinks due to interactions with neutral particles or recombination events. Hence combining (7) with a Bohm-Chodura sheath criterion [145-147] $q_\parallel^{tgt} = \gamma \Gamma_\parallel^{tgt} T^{tgt} = [\gamma_e + \frac{5}{2} + (M_\parallel^{tgt})^2] M_\parallel^{tgt} n^{tgt} c_s^{tgt} T^{tgt}$, where $\gamma_e \approx 4.8$ is the electron sheath transmission factor [146,147], $c_s^{tgt} = \sqrt{(2T^{tgt}/m_i)}$ the local isothermal sound speed and $n_i^{tgt} = n_e^{tgt}$, $T_i^{tgt} = T_e^{tgt}$ have been assumed, then :-

$$\Gamma_\parallel^{tgt} \approx \varphi_p^2 \frac{(p^{mid})^2}{2m_i} \frac{1}{q_\parallel^{tgt}} \left( \gamma_e + \frac{5}{2} + (M_\parallel^{tgt})^2 \right) \left( \frac{M_\parallel^{tgt}}{1+(M_\parallel^{tgt})^2} \right)^2 \quad . \tag{8}$$

Equation (8) thus makes clear that for $M_\parallel^{tgt} \geq 1$, along-field ion flux density at the target will tend to rise as power density arriving is lowered by cooling due to an extrinsic radiator. For $\Gamma_\parallel^{tgt}$ conversely to fall requires non-zero momentum sinks $\varphi_p < 1$ too, or stated alternatively, the definitive signature for onset of detachment of target ion flux passing a maximum and beginning to decline for fixed or increasing upstream density [75] unequivocally signifies the advent of edge momentum losses.

This behaviour is examined for ILW pulses in Fig. 13 (left), where total ion flux impinging on the whole toroidal outboard target between ELMs has been estimated by integrating over embedded probe signals, analogously to (6), again averaged during 50 - 90% segments of inter-ELM periods within 1 - 2 s flat-top windows. Normalising to D fuelling rate then approximates the outboard flux-amplification ratio in steady state $\Im_{out}^{i-E}$, shown as a function of N seeding input. Recall subsets with best (plain symbols) and some missing (faint symbols) probe data are discriminated since the latter effectively represent lower bounds only. Flux amplification first rises as N is injected, establishing a moderately enhanced recycling regime, consistent with the gain in plasma density seen in Fig. 3 but also possibly reflecting a contribution to $\Gamma_{i\,h}^{tgt}$ of nitrogen ions themselves. As N seeding is increased further, within the uncertainty due to some incomplete probe measurements, $\Im_{out}^{i-E}$ then appears to reach a maximum and roll over, particularly at medium and lower fuelling. Hence as described in (8), onset of plasma detachment from the target is suggested, although it should be borne in mind its emergence is in response to injection of an extrinsic rather than the majority species. In other words, N seeding can promote access to a detached regime. A further check is presented in the middle graph of Fig. 13, where the same ordinate is replotted against the fractional heat load landing on the outboard target between ELMs, deduced from probes as in Fig. 11. Now it can be seen that while $P_{out.tgt}^{i-E}/P_{in}$ drops due to seeding, $\Im_{out}^{i-E}$ tends to increase as outlined in (8), before falling off for highest N input even as target power decreases still further. This roll-over seems most prominent for the pulse at lowest fuelling and highest seeding, which it will be recalled from above was also the instance with most pronounced reduction of $P_{out.tgt}^{i-E}/P_{in}$ in Fig. 11. Thus plasma momentum sinks do seem to be encouraged at higher N seeding, which can thereby assist onset of detachment.

It is not clear a priori whether dispersal of SOL momentum to surrounding walls grows owing to cooling by extrinsic radiation leading to more favourable conditions for scattering by majority atoms and molecules, or whether losses occur directly through collisions with N particles themselves. Detachment develops first around the peak in target ion flux density close to the magnetic strike-point, due to concentration of recycling and neutral-particle interactions in its vicinity. Although not explicitly measured, a consistent approximation of electron pressure at this location may be extracted from probe data according to :-

$$p_e^{tgt} \approx \Gamma_\parallel^{tgt} \sqrt{(m_i T_e^{tgt})/2} \quad , \tag{9}$$

where an ion flux consisting entirely of deuterons and $T_i^{tgt} = T_e^{tgt}$ have again been assumed, plus [145-147] $M_\parallel^{tgt} = 1$. Since the peak value of (9) between ELMs across the outboard target is picked out, attention has been restricted just to those pulses with best probe data more reliably defining such a maximum. Accompanying measurements of upstream SOL pressure were not available in these experiments, so an indicative normalisation has instead been invoked by dividing by the inter-ELM electron pressure pedestal height, derived from HRTS signals in Fig. 5; the resulting ratio would then relate to scalar pressure drop along the SOL if upstream separatrix and pedestal quantities varied in reasonably close proportion. Under this assumption, peak $p_e^{\text{out.tgt i-E}} / p_e^{\text{ped i-E}}$ is lastly plotted versus inter-ELM loss power fraction $f_{\text{loss}}^{\text{i-E}}$ in the right-hand pane of Fig. 13, revealing that within appreciable uncertainties, decay of the pressure ratio, i.e. increase of edge momentum sinks, is roughly in proportion to the cooling and ELM-cycle effects of N seeding. Note from (7) that a sufficiently supersonic flow at the target-sheath edge could also lower the SOL pressure ratio but from (8) would not reduce ion flux, as actually seen. Although not an unambiguous demonstration, therefore, the linear correlation in Fig. 13 (right) would tend to be consistent with momentum removal by reinforcement of majority interactions through the impact of impurity seeding on the diverted plasma. Detailed 2-D modelling of coupled edge plasma – neutral-particle transport is continuing to analyse the situation more quantitatively.

## 9. Divertor W sources

While metallic walls may help to alleviate co-deposition retention of fuel species, a potential disadvantage is their more damaging effect upon plasma performance, owing to stronger radiation and dilution[72], if they lead to contamination of its confined volume. An equally important requirement on ILW operations is therefore to avoid significant infiltration especially of eroded W into the plasma core, the primary source of which may be expected to emanate from the divertor targets (some W-coated tiles for shine-through protection are also located in the main torus[58,59]). Introduction of extrinsic impurities to moderate power exhaust may either help in this respect, by sufficiently reducing target-facing plasma temperatures to fall below sputtering thresholds, or else exacerbate W release due to the greater sputtering power of heavier, more highly-charged particles[70,71]. As observed in the Introduction, dominance of the former over the latter effect becomes the prima facie preferred operating condition.

Sources from the outboard side of the ILW divertor, including its bulk-W target plate, may be estimated from absolutely-calibrated visible spectroscopy sampling a wide-angle fan from the top of the machine at a rate of $\nu_{\text{spect}} = 25\,\text{Hz}$ in light from neutral WI line-emission at 401 nm. An important feature is removal of background light from the signal to obtain a reliably pure line intensity. The range of W atoms is likely to be very short owing both to their high mass and low ionisation potential[73] of 7.86 eV (see below), so selection of this wavelength ensures detection of actual W fluxes essentially as they leave eroded surfaces. To evaluate ionisation to photon emissivity ($S/(XB)$) coefficients, proximate electron temperature has first been determined between ELMs from embedded probe data, again averaging $T_e^{\text{tgt i-E}}$ over 50-90% portions of inter-ELM periods during 1-2 s flat-top windows, following (1). The strong dependence of $S/(XB)$ on $T_e$ implies least light per particle will issue from the target region of peak temperature, or stated conversely, using its peak value will set an upper bound on the inferred W flux. On the other hand, using an average value of $T_e^{\text{tgt i-E}}$ will provide a more moderate, global figure for W release rate. Temperatures both at the peak and an average over the whole toroidal area of the outboard target $\int R^{\text{tgt}} T_e^{\text{tgt i-E}} ds / \int R^{\text{tgt}} ds$ have therefore been extracted, as shown against N seeding rate in Fig. 14 (left), restricting attention to those pulses with probe data best defining target profiles. These results confirm that near-target electron temperature between ELMs falls by at least a factor of 2 under N seeding and even at the peak approaches a time-average of ~10 eV or less, below which W sputtering yields decline rapidly[70,71].

Owing to the relatively slow sampling rate and correspondingly long period $1/\nu_{\text{spect}} = 40\,\text{ms}$ over which each spectrometer measurement is averaged, a slightly less strict criterion for choosing its inter-ELM points has been applied to increase statistics, viz. data are accepted where they fall 25‑90% of the way between successive ELM peaks and are simultaneously at least $1/(2\nu_{\text{spect}})$ away from either peak. After release, an atom of W can linger in the divertor plasma only for a time of order $\tau_I = 1/\{n_e^{\text{tgt}}\langle\sigma v(T_e^{\text{tgt}})\rangle_I\}$, where $\langle\sigma v\rangle_I$ is the Maxwellian rate coefficient for electron-impact ionisation[164], before being ionised. As hinted above, even for modest properties $n_e^{\text{tgt}} \sim 10^{17}\,\text{m}^{-3}$, $T_e^{\text{tgt}} \sim 5\,\text{eV}$ and ground-state atoms, this duration is just $\tau_I \approx 50\,\mu\text{s}$, so WI (401 nm) radiances $\Re$ thus sorted can indeed be taken to represent W fluxes evolved predominantly between ELMs. The small inconsistency with $T_e^{\text{tgt i-E}}$ selection is expected to be well within other uncertainties. An illustration of a resulting sample over the same (1.75 s) flat-top window is shown for an unseeded pulse in the top-right inset of Fig. 14, where identified points are depicted in red and their associated detection intervals are shaded in pink. This inter-ELM subset in each plasma is finally averaged analogously to (3), to preserve asymmetric variances, and combined with peak or toroidally-averaged $T_e^{\text{tgt i-E}}$ using a recent multi-machine determination[165,166] of $S/(XB)$ to interpret a flux density of W atoms, viz. :-

$$\Gamma_W = \hat{\Gamma}_W / A_{\text{surf}} \approx 4\pi\Re(S/XB) \quad (\text{m}^{-2}\cdot\text{s}^{-1}) \;, \qquad (10)$$

where uncertain area of the emitting surfaces $A_{\text{surf}}$ has been deliberately suppressed. Respective quantities are plotted as functions of N input in the centre panes of Fig. 14.

Variation of $\Gamma_W^{i-E}$ is qualitatively the same whether considering peak or areally-averaged $T_e^{\text{tgt i-E}}$ and actually reveals that, under seeding, WI emisson between ELMs is often close to or at the detection limit of the spectrometer. Nevertheless, within appreciable uncertainties, a decline of $\Gamma_W^{i-E}$ with increasing N puffing rate is suggested, consistent with the fall in inter-ELM near-target temperature and outboard-target heat load, Fig. 11. It also complies with a previous finding for ILW N-seeded L-mode plasmas[166], similarly deducing a drop in target erosion when $T_e^{\text{tgt i-E}}$ was lowered below $\approx 15\,\text{eV}$. Consequently it is hinted that inter-ELM erosion of W primarily by intrinsic Be in unseeded pulses[166,167] may be reduced by cooling due to an extrinsic radiator, and seeding sufficient to achieve this effect also does not tend to cancel the benefit by sputtering due to N itself. In other words, its impact appears to be encroaching on the condition advocated as preferable for operations above.

However, separate spectroscopic studies[166,167] have indicated that net W erosion in Type I H-mode is typically dominated by that occurring at the ELMs themselves. Furthermore, as divulged in Figs. 7I & 11, injecting N into ILW plasmas tends to lower frequency of their ELMs, with a commensurate increase in their energy amplitudes, so these more severe transients would be apt to realise still greater sputtering power. To obtain a preliminary estimate of neutral W fluxes at such events, spectrometer data have additionally been filtered to select those points falling within $1/(4\nu_{\text{spect}})$ of an ELM peak, which should therefore be dominated by its associated emission. An example of this scheme is again included in the top-right inset in Fig. 14, where the qualifying points are marked in green and their detection intervals shaded in yellow. Averaging the ELM subset of WI radiance measurements in the 1‑2 s flat-top window of each pulse analogously to (3), approximate flux densities have then been inferred from (10) assuming a constant $(S/(XB))^{\text{ELM}} \approx 50$, representing the saturated level approached for high electron temperatures[165,166] likely to be reached during rapid Type I excursions, but not well tracked by the Langmuir probes; as above, this treatment can also be taken to express an upper bound on $\Gamma_W^{\text{ELM}}$. Final momentary W flux densities are plotted for the same group of plasmas in the bottom right pane of Fig. 14. Although the analysis is crude, a rise of $\Gamma_W^{\text{ELM}}$ with N input is first clearly implied for those instances where ELM frequency falls and their amplitude increases (cf. Figs. 7I & 11), before it drops again as more moderate fluctuations are reasserted at higher seeding. Strongest cases are even modestly reduced compared to unseeded ones, in line with smaller ELM sizes displayed in Fig. 11. Recall these transitory values do not immediately convey the effective W source rate, which is more properly related to $\sim \Gamma_W^{ELM} \cdot \Delta t^{ELM} \cdot \nu_{ELM}$, where $\Delta t^{ELM}$ is a requisite time-scale of bright ELM emission. Hence Figs. 14 & 7I suggest that W atomic fluxes overall can be mitigated between and perhaps to a lesser extent at ELMs by N injection, while exhibiting no evidence of significantly aggravated erosion due to extrinsic N ions for the conditions addressed.

The typically short range of W atoms already mentioned means that eroded fluxes may be expected to be further attenuated by prompt recapture of ions born on Larmor orbits directly reintersecting the target surface [166]. On the other hand, their actual effect depends finally not just upon their magnitude but upon their transport into the confined plasma and concentrations thereby reached over longer time-scales. At least with respect to plasma purity, arbitrarily large divertor sources of W could be tolerated if they were kept well contained within its volume and only led to acceptably low core contamination, for example, whereas conversely even minor ELM fluxes may tend to be more penetrating through the SOL plasma owing to their more intense form $\Delta t^{ELM}/\tau_E \ll 1$. Diminished target sources would thus be a desirable but not sufficent condition to determine the core response to W, one aspect of which is considered next.

## 10. Non-stationarity

The impact of both extrinsic and intrinsic impurities on plasma performance depends ultimately upon the combination of their sources with their transport and confinement. The tendency for density to rise with N seeding in ILW H-modes has been described above (cf. Fig. 3), so in parallel with improving energy confinement it may be expected that confinement of particles is also increasing, particularly for very heavy species like W. Without adequate divertor retention and edge screening of even low target sources, therefore, core accumulation of W could be a potential hazard of N input. This is assessed indirectly in Fig. 15(a), where two measures of core radiation are compared for a pair of ILW pulses at higher fuelling, one without (#82751) and one with (#82811) N dosing. Time-traces of soft X-ray (SXR) and unfiltered total radiation are each plotted during flat-top windows along horizontal lines-of-sight through the core plasma; note that sawteeth are very well observed on the former, but ELMs are practically absent from the latter, suggesting their radiation amplitudes are not only smaller than in all-C counterparts (cf. Fig. 6) but also more local to the divertor region as well. In the unseeded case, both types of emission remain reasonably steady, but when N is added there is manifestly a monotonic increase of each signal, indicating a progressive change of core conditions. The growth of SXR intensity occurs despite relatively little change in line-averaged plasma density or $Z_{eff}$, so is not attributable to bremsstrahlung, instead being most likely to arise from line emissions due to accumulation of a sufficiently heavy impurity, i.e. W. This would be consistent with the simultaneous rise of total radiation too and implies the gain of inter-ELM radiated power fraction from inside the separatrix shown in Fig. 8(b) is probably a product of worsening W contamination over the fuelling/seeding scan, although not severely enough to degrade confinement. Derivation of actual W concentrations in the core from spectroscopic data is a challenging task still under development [168] but the typical behaviour of ILW seeded plasmas illustrated in Fig. 15(a) thus points towards a plasma W content which gradually builds up under N seeding.

An immediate consequence of deteriorating purity, evident in Fig. 15(a), is that N-seeded pulses so far in the ILW did not tend to achieve a comprehensively steady state. The extent to which their properties became unsteady due to the influence of W has been quantified by fitting an exponential form $a\exp\{t/\tau_{rise}\} + b$ to SXR and total-radiation time-traces along core lines-of-sight within 1-2 s flat-top windows, using a non-linear least-squares optimisation with Gaussian errors. Resulting rise times normalised to energy-confinement time averaged over the same intervals $\tau_{rise}/\tau_E$ are presented in Fig. 15(b) as functions (left) of N input and (right) of mean ELM frequency from Fig. 7(c). Corresponding all-C values (open symbols), which obviously were not affected by W, are also superimposed for comparison. In this representation, steadiest conditions are actually defined in unseeded ILW plasmas at highest fuelling (e.g. as in Fig. 15(a)), but they remain conspicuously more steady even with strong N seeding in all-C cases too. In contrast, any level of N input into ILW pulses at once substantially reduces their stationarity during flat-tops, irrespective of their accompanying fuelling rate. Note in this context that ILW plasmas at lowest fuelling, identified as exhibiting highest electron pedestal temperatures in §3.3, are always amongst the least steady instances even when unseeded. By examining core radiation, the effect is to integrate over impurities originating both from between and at ELMs, which are then liable to be smoothed out by transport. Perhaps partly in consequence, replotting the same ordinates against mean ELM frequency in Fig. 15(b) indicates no ILW correlation of $\tau_{rise}/\tau_E$, but there is at least no evidence of unstationary properties being related to larger, less frequent ELMs induced by N in the metallic environment. In other words, impurity accumulation does not seem to be controlled by higher frequency ELMs, in contrast to what has been inferred elsewhere for its unseeded H-modes[85,86]. The more general sensitivity to N injection into ILW plasmas denoted would be consistent with preceding inference of persistent time-average W sources from ELMs, or any reduction in W escaping from the divertor being offset by higher confinement of impurities still polluting the core. Note that any changing W sources outside the divertor have also not been ruled out.

All-C and ILW experiments here are not thoroughly equivalent since it will be recalled from §2 that the former incorporated central ICRF heating to help regularise their sawteeth[96], whereas this extra power was absent from the latter. It is evident in Fig. 15(a) that for the N-seeded pulse (#82811) sawtooth oscillations gradually diminish in amplitude as core emission rises and eventually become negligible after time $t \approx 17$ s, whereupon the rate of increase of total radiation accelerates considerably. An accompanying adjustment of the core current density profile is therefore also suggested, which was in fact typical of ILW pulses with N input and which could similarly have affected their tendency to become unsteady through a build up of radiated power due probably to W accumulation. Whether central ICRH can sustain sawteeth and so help to remedy this major deficiency of H-modes so far in the ILW will be tested in future campaigns.

## 11. Summary and conclusions

Installation of the ITER-like wall (ILW) on JET, to address the impact of its Be main torus and W divertor on next-step-relevant plasmas, has presented a unique opportunity to compare behaviour with that in its former all-C lining. In particular, such comparisons have been explored for Type I H-modes seeded with extrinsic impurities, ultimately necessary to obtain tolerable exhaust loads in both ITER and JET at full power and pulse length[69]. Systematic scans of D fuelling and N seeding have been matched in both JET designs, concentrating upon high-triangularity, single-null diverted plasmas at 2.5 MA, $q_{95} \approx 3.5$, which in the all-C machine exhibited most robust confinement against strong fuelling to high density[77-79]. Contrasting results first emerged for unseeded pulses in the ILW, however, which did not reproduce this regime and, even for the same configuration, input power, fuelling and measured torus gas pressure, instead yielded $\approx 10\%$ lower density and $\approx 30\%$ lower normalised confinement owing to a correspondingly cooler (electron) temperature pedestal. Incorporating N seeding then also elicited a different response, with density rising and ELM frequency initially falling to define a minimum as N input increased, both opposite to all-C effects. In addition plasma purity inferred from visible bremsstrahlung was significantly improved in the ILW, as expected for all-metal walls, while radiation, at first intrinsically lower, could be raised smoothly with N up to a power fraction of ≈0.6 between ELMs, again unlike the non-monotonic variation in all-C cases. In both contexts, this led as intended to a reduction in divertor inter-ELM heat load, estimated at the outboard side from embedded Langmuir probes or fast infra-red thermography, which, within temporal uncertainties, fell in proportion to the combined radiative cooling and effect upon ELM recurrence due to injected N. Energy amplitudes of ELMs were decreased when their frequency was increased in both environments too, although net power expelled by these transients generally still exceeded that exhausted in between them. More surprisingly, the latter average power exiting in ELMs itself tended to decline in proportion to the moderating impact of N in the ILW, departing from the uncorrelated alteration of all-C ELM power. Onset of detachment at the outboard target was simultaneously promoted by N input, the drop of approximate edge pressure ratio once more in proportion to N influence in the ILW hinting that this might be due to encouragement of plasma momentum dispersal by majority interactions.

The most striking contrast between all-C and ILW H-modes with N seeding concerned the reaction of the electron pedestal, detected by high-resolution Thomson scattering (HRTS). Whereas in the all-C environment this always tended to be degraded by extrinsic impurities, in the ILW its lower unseeded plateaux between ELMs could actually be raised by N, even eliciting a gain in pedestal temperature and pressure back to the levels seen in equivalently seeded all-C pulses. A similar effect had been reported in the all-W ASDEX Upgrade machine [25-27] but in these experiments has been seen on JET for the first time [88]. Normalised confinement then recovered directly in proportion to the restored pedestal, even while radiation from inside the separatrix itself increased, with electron transport in the core plasma remaining approximately unchanged. At the same time, altered ELM time-scales, frequency response to N seeding and departure from simple Fishpool scaling [119] of confinement as it varied all point towards different ILW pedestal dynamics and an effective improvement in local stability due to added N. In parallel, normalised ELM sizes did not display conventional inverse correlation with electron collisionality at the pedestal top [141] in either JET variant.

A fundamental effect of switching from an all-C to an all-metal first-wall is therefore the pronounced change in H-mode pedestal properties. In part this may be related to their markedly different recycling characteristics, and hence boundary conditions imposed upon their respective plasmas, as evidenced in the ILW divertor by inboard-outboard asymmetric transients at ELMs and emergence of spontaneous oscillations between them in unseeded states. The tendency of N seeding to recover (not surpass) the behaviour of all-C references, though, suggests that the degraded pedestal in unseeded ILW pulses may be due to the absence of intrinsic C, the missing effects of which can then be substituted by extrinsic N. A hitherto unrecognised role of (light) impurities in pedestal stability [88] would thus be implied, whereby its height and so global confinement would be improved owing to benefits of their greater inertia, or charge, or both, than majority ions. The underlying loss of confinement in unseeded ILW plasmas would then be associated with their edge being effectively too pure. Further theoretical and experimental studies are required to test this hypothesis rigorously.

These many differences between equivalent H-modes in the alternative JET wall environments demonstrate the profound influence which boundary conditions and intrinsic impurities can exert upon their performance. Sources of intrinsic W in the ILW divertor tended to decrease between ELMs with N seeding, though perhaps weaker net attenuation of those at ELMs, together with rising particle as well as energy confinement, still led to persistent core contamination and substantially less stationary conditions than in all-C pulses. Unlike some previous unseeded studies [85,86], there was no indication this accumulation could be mitigated by higher-frequency ELMs. Such unsteadiness remains the major deficiency of ILW "baseline" operation and constitutes a high priority for forthcoming campaigns, first by testing whether central ICRH can alleviate the problem by helping to sustain sawteeth, as applied in all-C instances. Furthermore, real-time control of extrinsic seeding will be pursued in order to extend better stationarity to pulse durations eventually up to $\approx 20$ s, as well as to seek stricter control of divertor heat load via stable imposition of higher radiation fractions between ELMs, closer to the level of $\approx 85\%$ appropriate for ITER [11,12]. While target power load was successfully reduced by added N both between and even for ELMs in the ILW, other developments required include combination finally with active ELM moderation in order to limit the size of individual bursts, particularly as total heating is gradually raised towards a maximum of $\approx 34$ MW [69]. In addition, it is vital to adapt the seeding strategy to a noble gas such as Ne, since N will not be compatible with active-gas handling in future D-T experiments on JET and, due to ammonia formation, may also be disfavoured on ITER. Primary issues, for example, will be how to accommodate recycling properties of Ne expected to be different again from those of N, how to control W sputtering of even more erosive Ne ions, and above all whether the beneficial recovery of higher pedestal pressure with N can be reproduced with Ne. Together such developments should continue the progress reported towards a fully-integrated, impurity-seeded H-mode scenario suitable for exploitation in high-power, long-pulse D-T plasmas on JET and thereafter during the inductive phase on ITER.

*This work, part-funded by the European Communities under the contract of Association between EURATOM / CCFE, was carried out within the framework of the European Fusion Development Agreement. For further information on the contents of this paper please contact publications-officer@jet.efda.org. The views and opinions expressed herein do not necessarily reflect those of the European Commission. This work was also part-funded by the RCUK Energy Programme [grant number EP/I501045].*

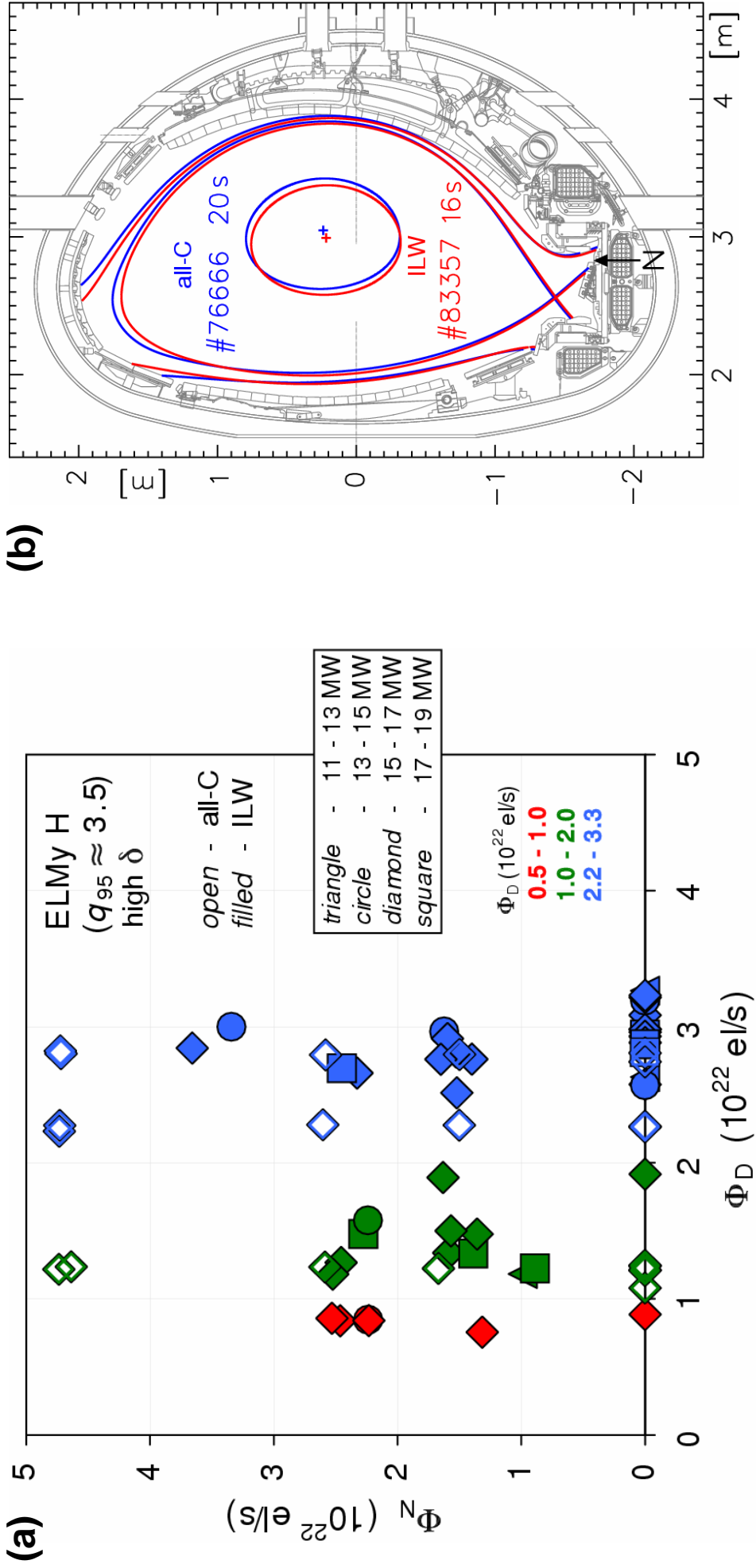

Fig. 1 (a) Deuterium-fuelling plus nitrogen-seeding scans conducted in high-triangularity H-modes at 2.5 MA in the all-carbon (*open symbols*) and ITER-like-wall (ILW) (*filled symbols*) designs of JET. Inputs are specified as time-averages of electrons added per second during pulse flat-tops, assuming full ionisation. Points are colour-coded according to the scheme used to indicate low, medium and high fuelling levels in subsequent scatter plots. In this and the following figure only, different symbols also designate total heating power involved. Similar ranges are covered in both environments, with most pulses between 15 - 17 MW. (b) Magnetic equilibria from EFIT[89-91] reconstruction in typical all-C and ILW flat-tops at ≈16.5 MW, showing closeness of the replication. The scrape-off-layer (SOL) surface drawn is 4 cm outside the separatrix at the outboard centre-plane. *Superimposed*: inlet position of nitrogen seeding from the divertor outboard floor.

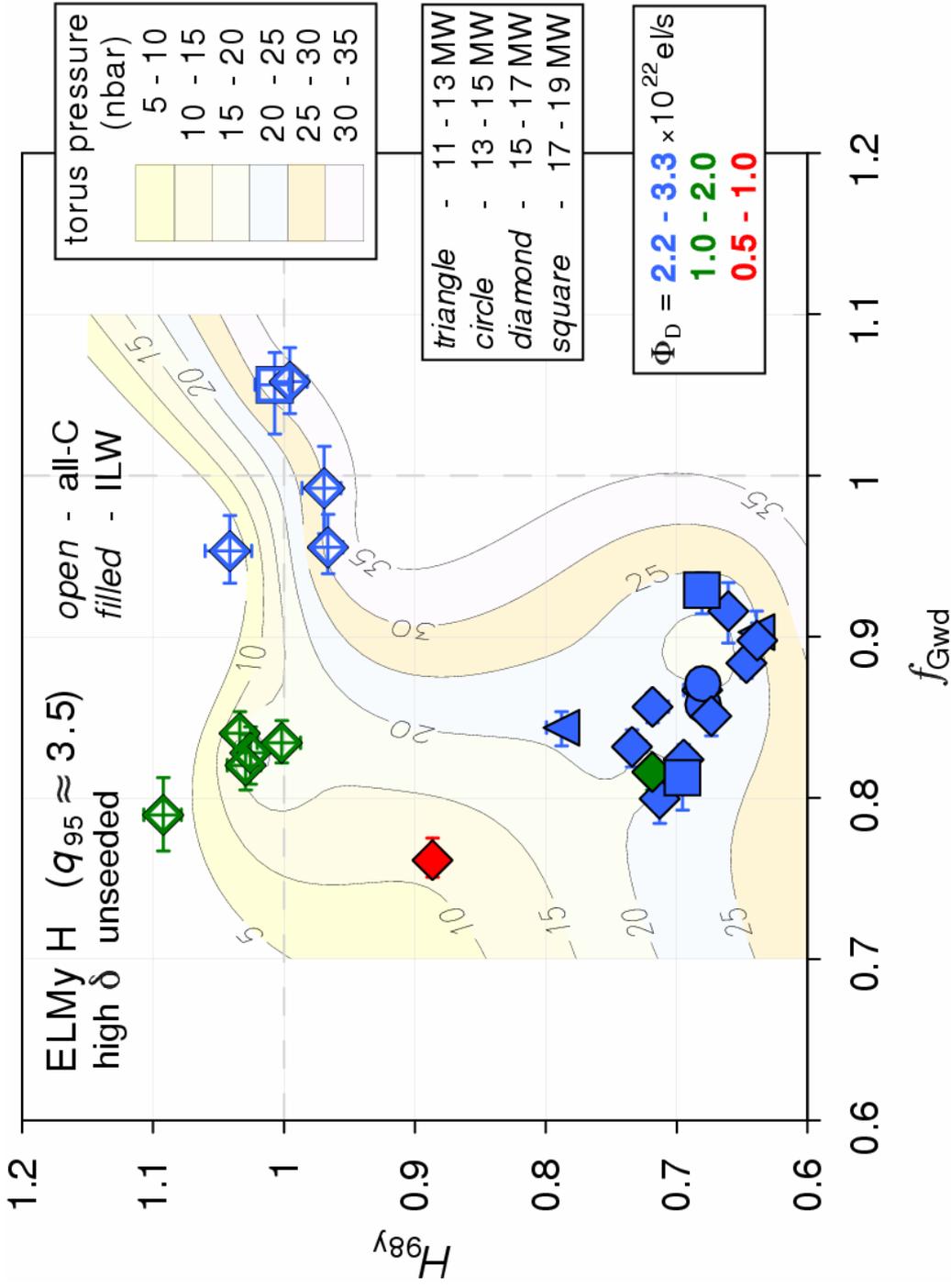

Fig. 2 Performance summary for JET unseeded high-triangularity ELMy H-modes at 2.5 MA, in terms of energy confinement normalised to the ITERH98(y,2) scaling law[102] versus line-average density normalised to the Greenwald value[66]. *Open symbols*: all-C machine; *filled symbols*: ILW. Points are colour-coded according to their level of deuterium fuelling and, in this and the preceding figure only, different symbols discriminate total input power levels. *Also superimposed*: contours of torus gas pressure in nanobars. Confinement is lower and declining with density in the ILW, despite similar shape, input power, fuelling rate and torus pressure to all-C counterparts. (NGE)

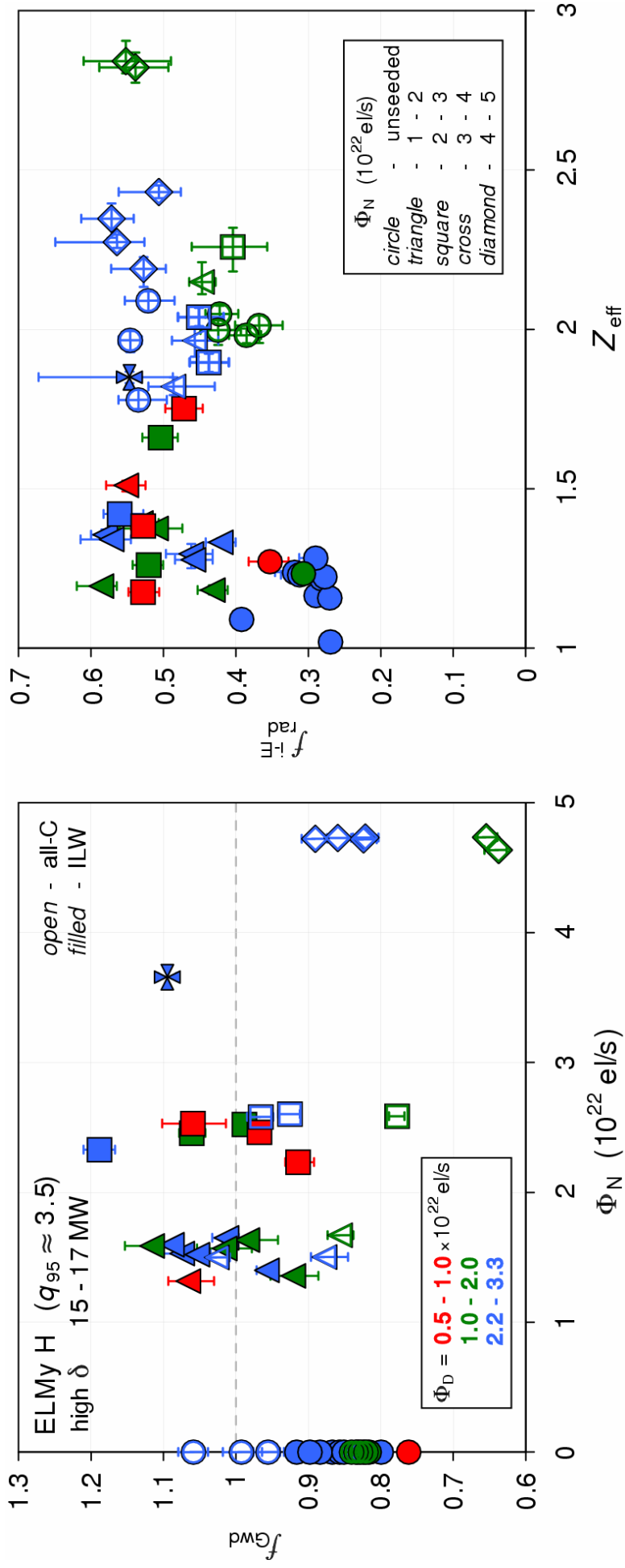

Fig. 3  Response of JET high-triangularity ELMy H-modes at 2.5 MA and total input power between 15 - 17 MW to nitrogen seeding into the divertor. *Left*: normalised density versus nitrogen input rate in terms of electrons added per second, assuming full ionisation. *Right*: total radiated power fraction, averaged between ELMs, versus line-average effective ionic charge from visible bremsstrahlung. *Open symbols*: all-C; *filled symbols*: ILW. Points are colour-coded according to their deuterium-fuelling rate, while here and in subsequent scatter plots, different symbols indicate nitrogen-seeding levels (note in particular circles are always unseeded). Density declines with seeding in the all-C cases, but rises with it in the ILW. Purity tends to be higher in the ILW and intrinsic radiation lower, while N seeding raises total radiation fraction between ELMs in either environment, but not above ≈0.6. (NGE)

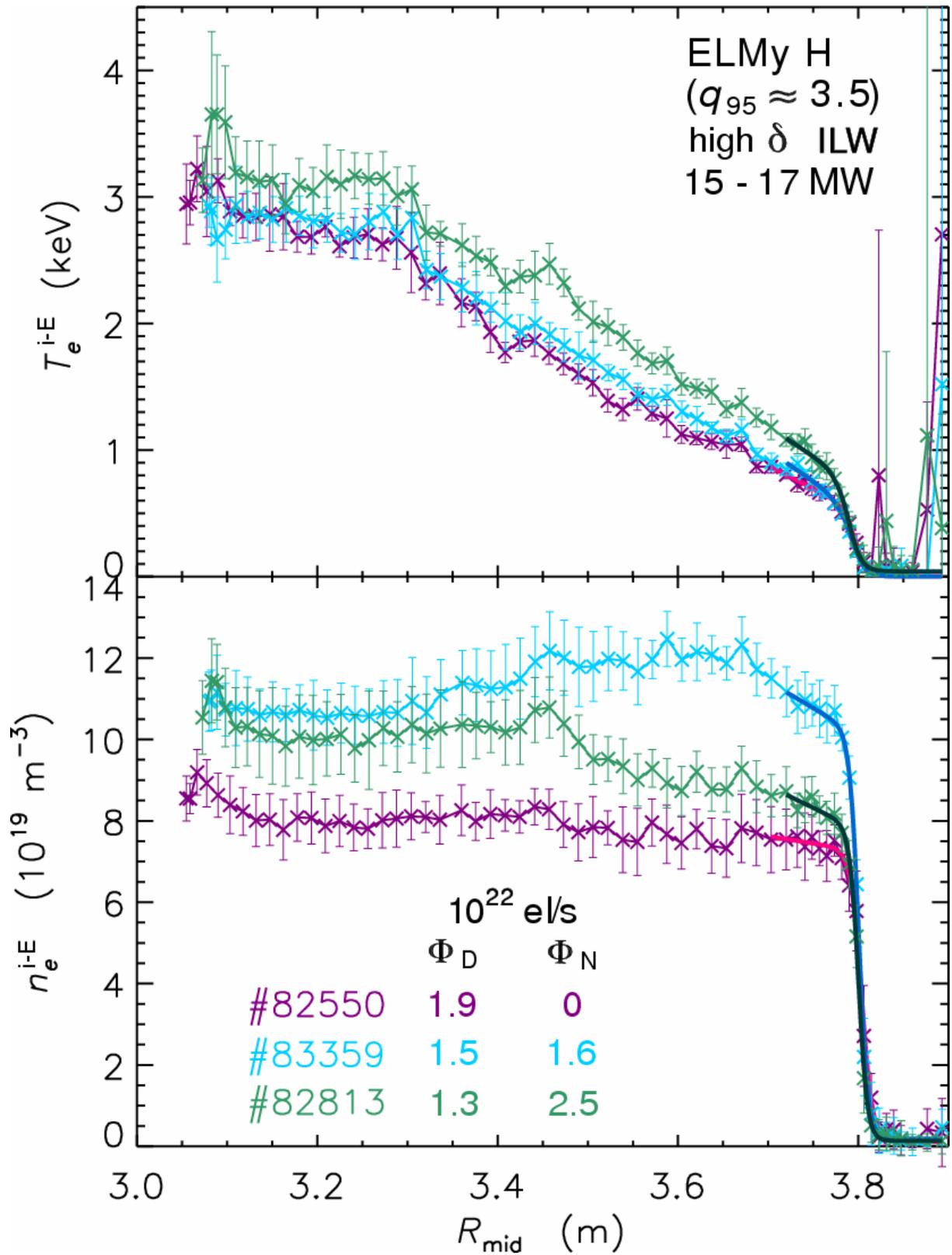

Fig. 4(a)  Example electron temperature (*top*) and density (*bottom*) profiles versus mid-plane major radius, during the flat-top at high triangularity, 2.5 MA, 15-17 MW and intermediate fuelling in the ILW, from high-resolution Thomson scattering (HRTS)[103,104], averaged within the final third of inter-ELM periods. *Solid lines*: fitted modified tanh functions[126-129] for normalised minor radius $\rho \geq 0.85$. N seeding first mostly raises the pedestal density, but then increases the pedestal temperature for somewhat lower density again.

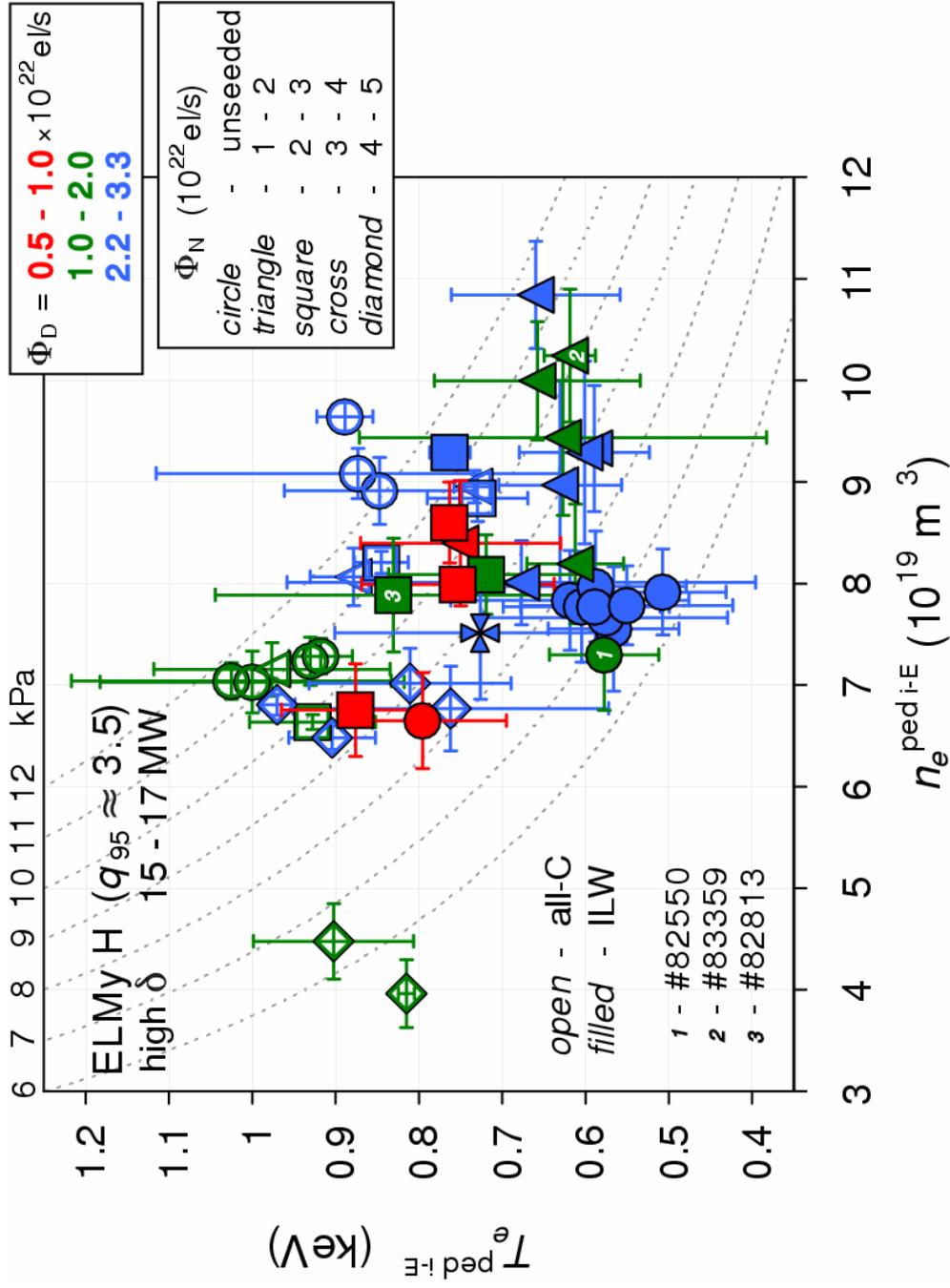

Fig. 4(b) Electron pedestal temperature versus density diagram for high-triangularity ELMy H-modes at 2.5 MA with total power between 15-17 MW. Each point is obtained from fitting modified tanh functions [126-129] to HRTS profiles ($\rho \geq 0.85$) averaged in the final third of intervals between ELMs during pulse flat-tops, with colour and symbol coding of D and N inputs respectively (as in Fig. 3). *Open symbols*: all-C; *filled symbols*: ILW. The three ILW example shots in Fig. 4(a) are specifically labelled. N seeding in the all-C machine tends to lower the pedestal, mostly by decreasing the density. Unseeded pedestals in the ILW tend to be cooler, but N seeding first raises the density, then the temperature, for pressure recovered to corresponding all-C levels.

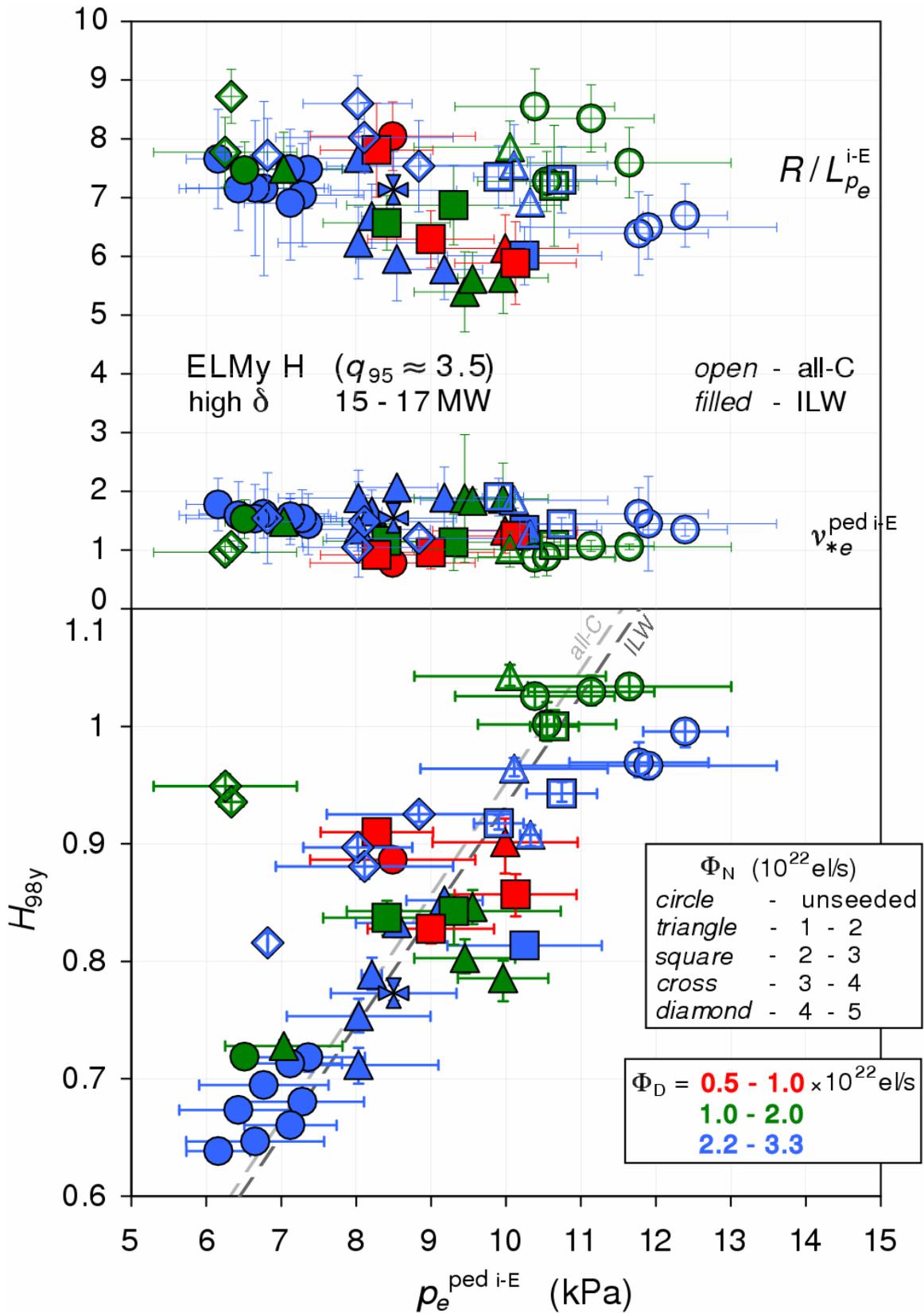

Fig. 5  Properties of high-triangularity H-modes at 2.5 MA and power 15 - 17 MW against electron pedestal pressure. *Top*: normalised inverse scale-length of electron pressure over $0.3 \leq \rho \leq 0.8$ in normalised minor radius; electron collisionality[132] at the pedestal top (error bars shown faint for clarity). *Bottom*: normalised global energy confinement[102], *superimposed*: best linear fits to data for each environment.

*Open symbols*: all-C; *filled symbols*: ILW. Points colour and symbol coded for D and N inputs respectively (as in Fig. 3). Pedestal data are from modified tanh fits [126-129] to HRTS profiles ($\rho \geq 0.85$) averaged in the last third of inter-ELM periods during flat-tops, core scale-length from a linear fit to the same average profiles. Pressure peaking does not increase and pedestal collisionality varies little over the datasets, even though pedestal height and confinement are clearly falling with N seeding in the all-C cases and rising with it in the ILW. Similar values of these are eventually recovered in both situations, while almost the same direct proportionality is maintained throughout. ($H_{98y}$ NGE)

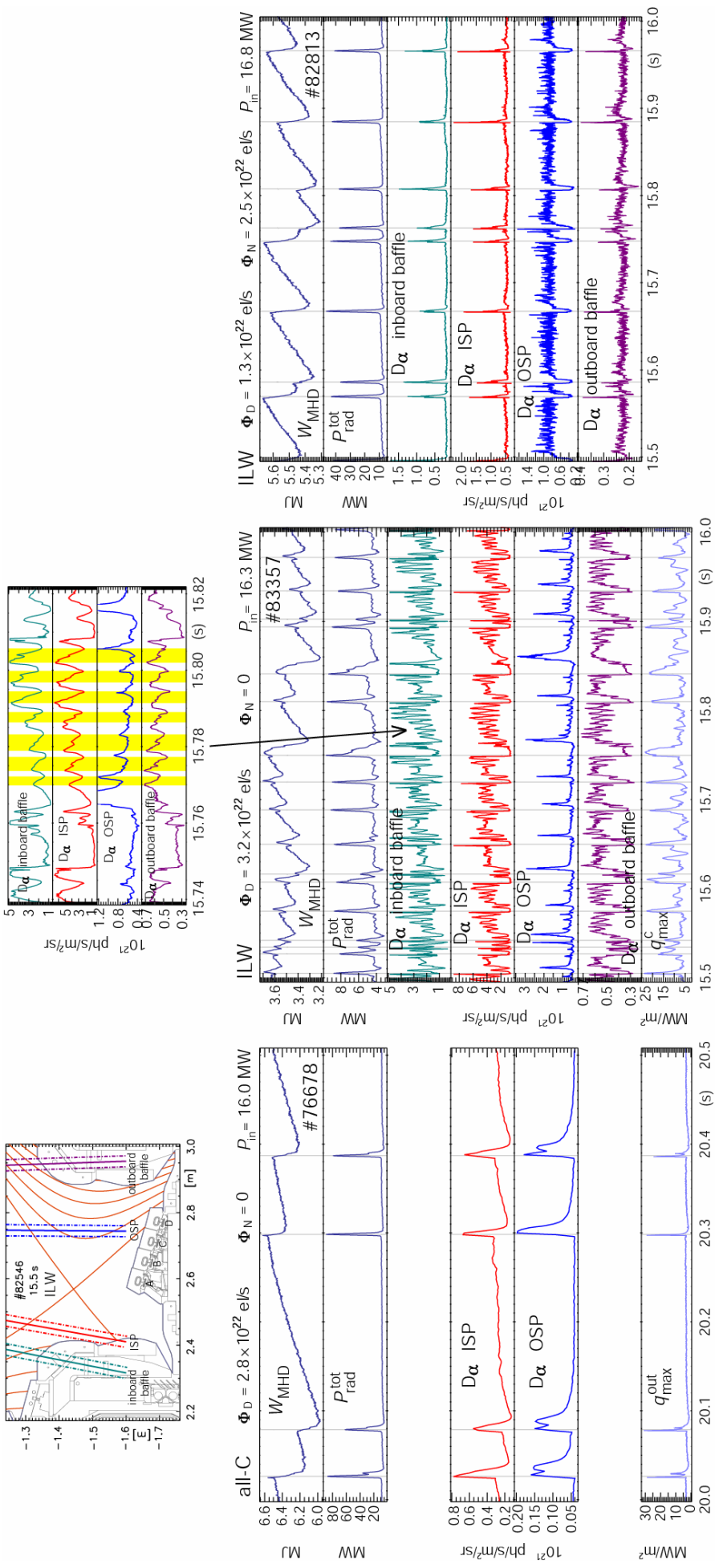

Fig. 6  Example time-traces during pulse flat-tops for high-triangularity H-modes at 2.5 MA and input power 16 - 17 MW. *Signals from top*: fast reconstruction of plasma stored energy; total radiated power from bolometry; Balmer-α (656 nm) radiance along vertical lines-of-sight viewing respectively the inboard baffle, inboard strike-point (ISP), outboard strike-point (OSP) and outboard baffle (*see inset top-left*); peak power loading on the divertor outboard target from fast infra-red thermography [133] (whole plate for all-C, stack C for ILW). *Left*: higher-fuelling, unseeded all-C; *centre*: higher-fuelling, unseeded ILW; *right*: intermediate-fuelling, higher-N-seeding ILW (note different vertical scales across each row). ELMs in every all-C case always present as fast, nearly simultaneous rises in radiation, recycling at the inboard and outboard strike-points and target heat load. In contrast for the ILW, they elicit a drop in ISP and a delayed rise in OSP recycling for unseeded pulses. This can reverse (ISP rise, simultaneous OSP drop) under N seeding. Unseeded ILW plasmas also tend to exhibit inter-ELM oscillations in divertor recycling which are in exact anti-phase between the ISP and OSP (*see inset top-centre*).

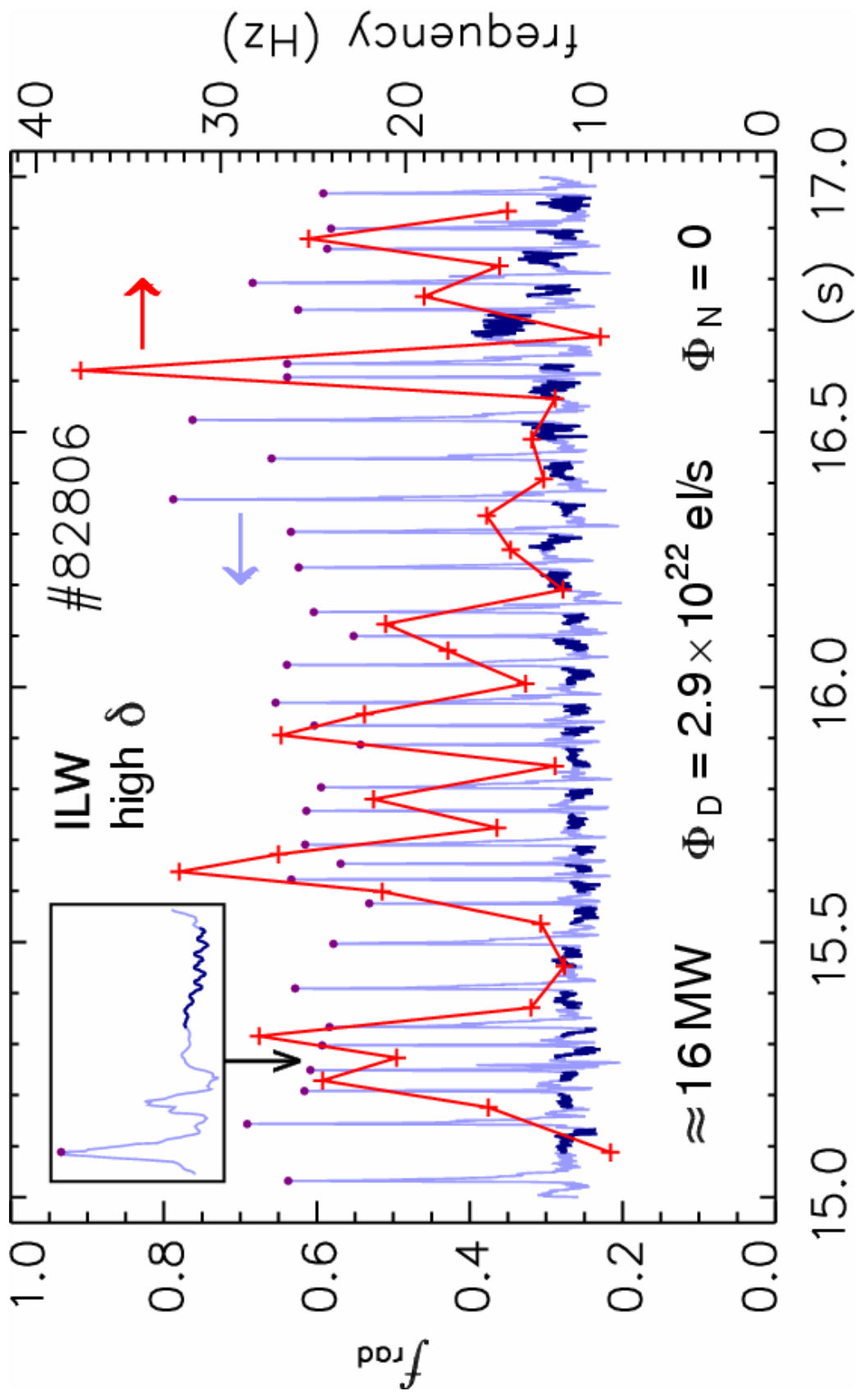

Fig. 7(a)  Example identification of ELMs from time-trace of total radiated power fraction (*left-hand ordinate*) during the flat-top of a high-triangularity, unseeded, 2.5 MA H-mode at higher fuelling and ≈16 MW in the ILW. Located ELM peaks labelled by dots, dark segments indicate intervals 50% - 90% of the way between successive peaks, used for inter-ELM averaging of properties (e.g. radiation, Langmuir probe data, etc.). The reciprocal of each whole period between consecutive peaks defines a momentary ELM "frequency" (*right-hand ordinate*). *Inset*: $f_{rad}$ around the ELM at 15.25 s on an expanded time scale.

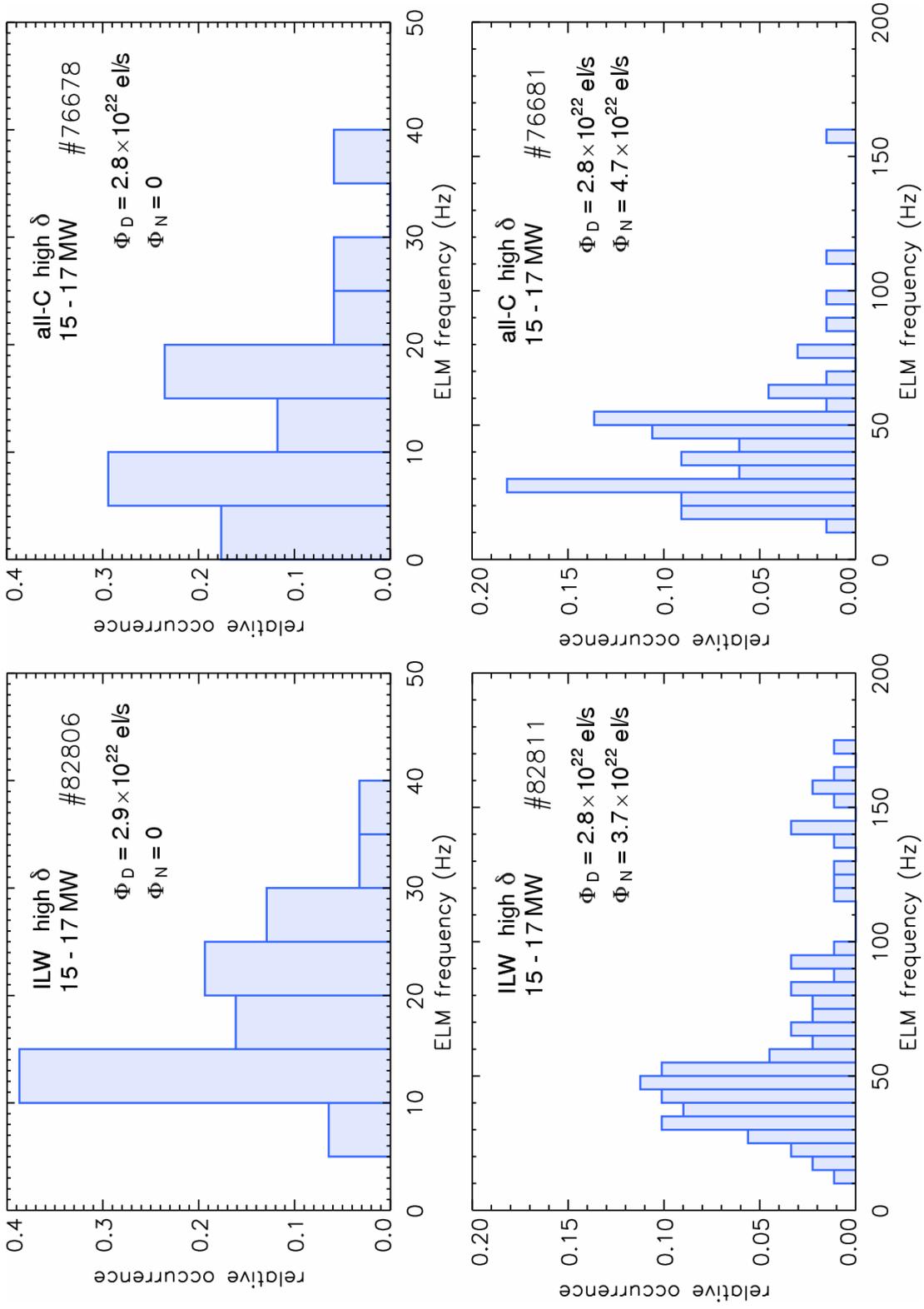

Fig. 7(b)  Example ELM frequency normalised distributions, accumulated with 5 Hz resolution, during the flat-tops of high-triangularity, 2.5 MA H-modes at higher fuelling and 15 - 17 MW. *Left*: ILW; *right*: all-C; *top*: unseeded; *bottom*: higher N seeding. Distributions are generally far from normal and tend to broaden under seeding.

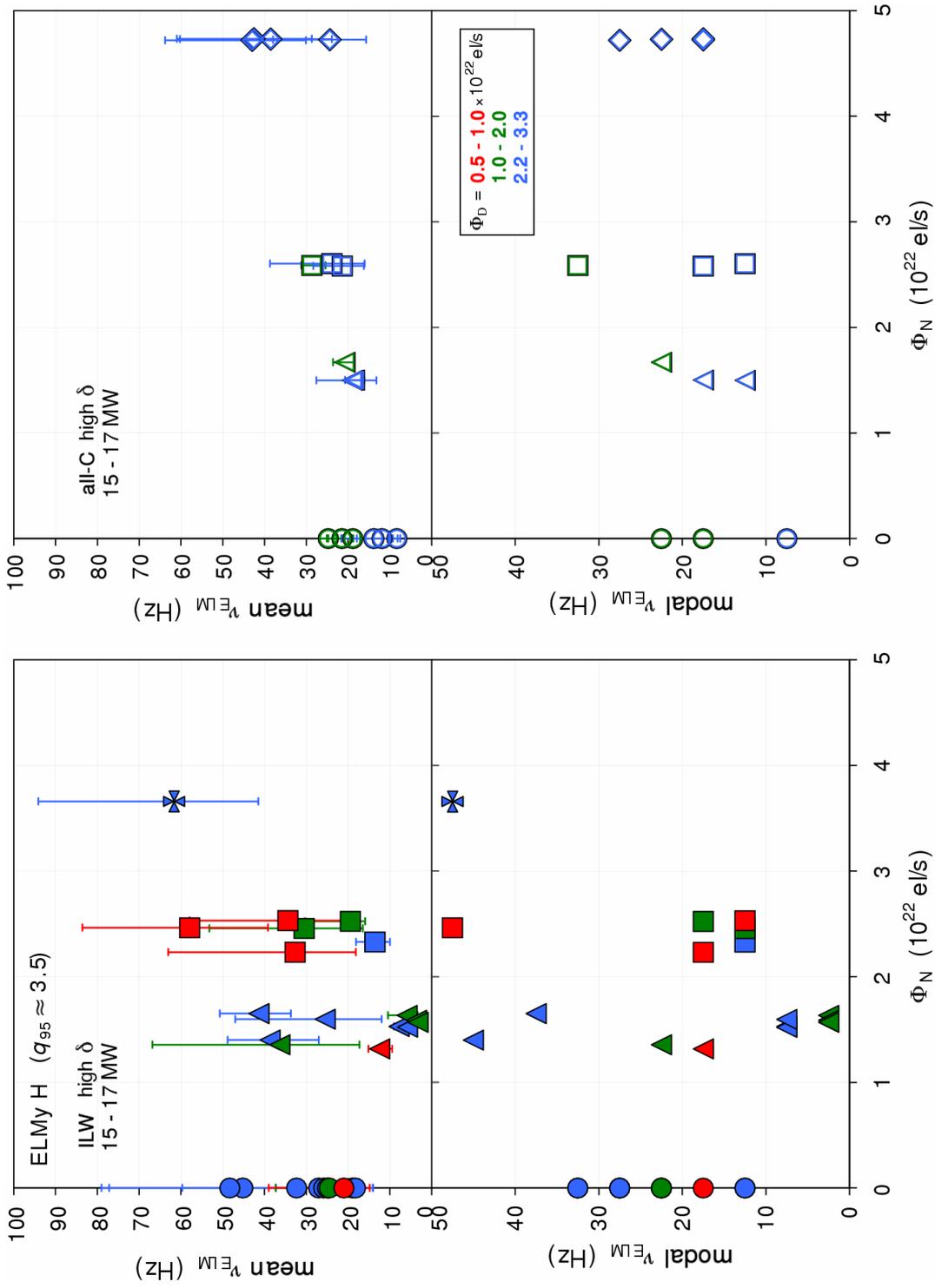

Fig. 7(c) Characterisations of ELM frequency response to seeding in high-triangularity H-modes at 2.5 MA and 15 - 17 MW, as functions of N injection (electrons per second assuming full ionisation). Points colour and symbol coded according to D and N inputs respectively (as in Fig. 3). *Left*: ILW; *right*: all-C; *top*: mean frequency (NGE); *bottom*: modal frequency (with a fixed resolution of 5 Hz). ELM frequency clearly rises with seeding in the all-C machine, but tends first to fall with it in the ILW, defining a minimum value.

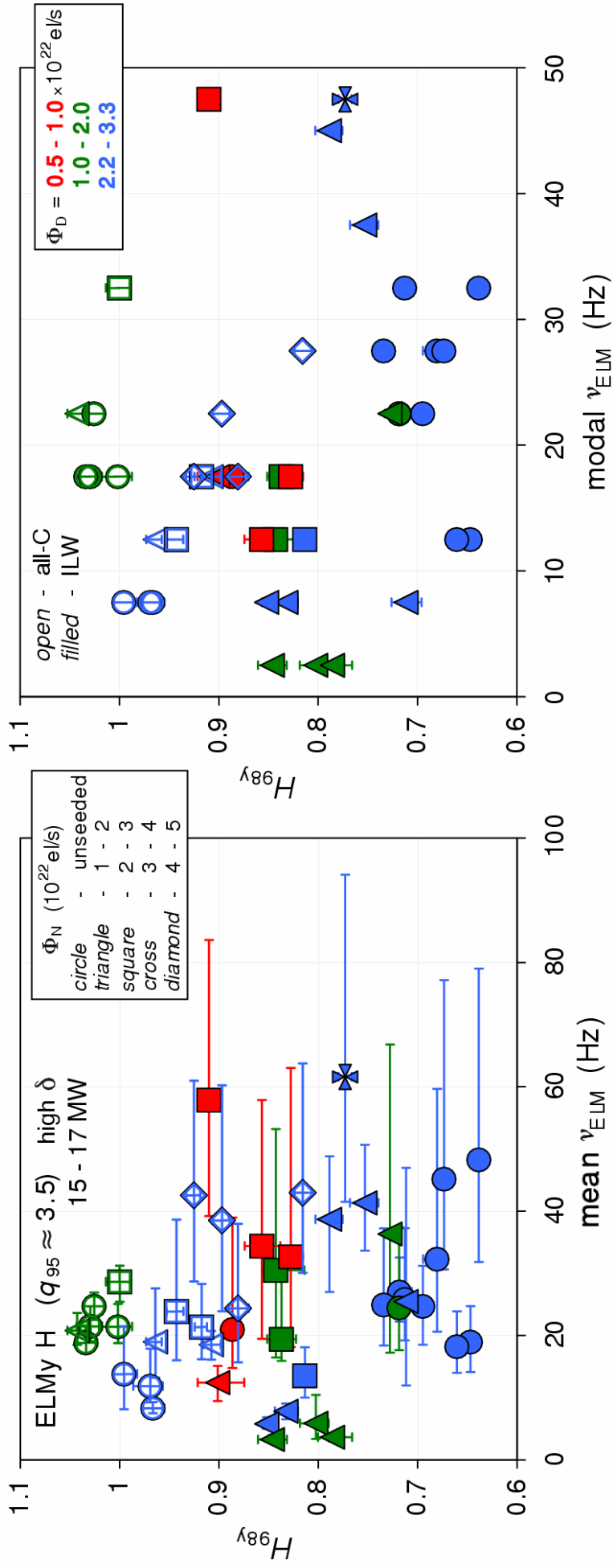

Fig. 8(a)  Normalised energy confinement in high-triangularity, 2.5 MA H-modes with power 15 - 17 MW as a function of mean (*left*) and modal (*right*) ELM frequency. *Open symbols*: all-C; *filled symbols*: ILW; points colour and symbol coded for D and N rates respectively (as in Fig.3). At highest fuelling, confinement is higher for lower ELM frequency in either environment, but this tendency becomes much weaker for best confined cases at lowest fuelling. (NGE)

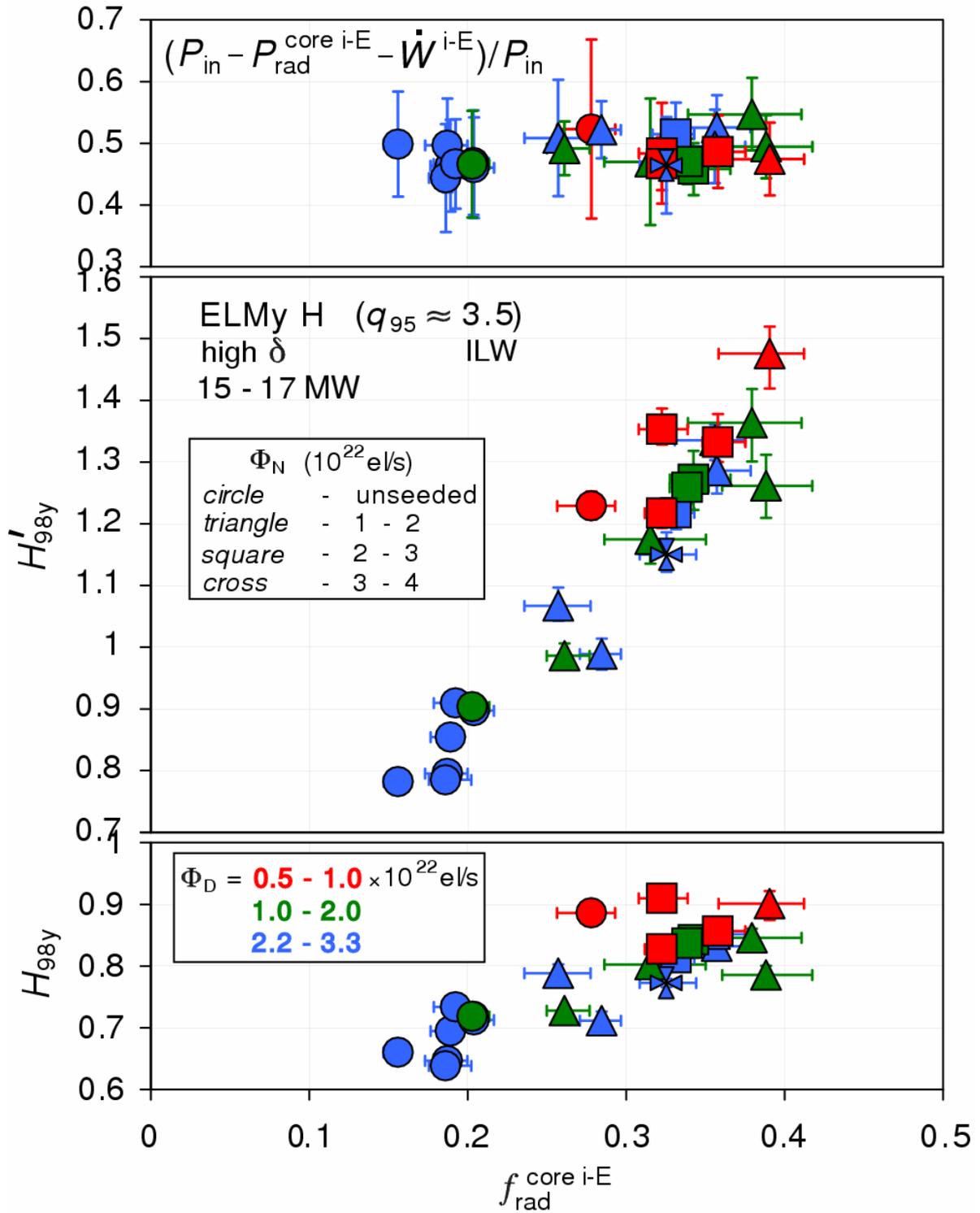

Fig. 8(b)  Energy confinement in high-triangularity, 2.5 MA H-modes at 15 - 17 MW in the ILW versus radiated power fraction from the confined plasma ("core") between ELMs, estimated from tomographic reconstruction of bolometer signals. Points colour and symbol coded for D and N inputs respectively (as in Fig. 3). *Bottom*: normalised to the ITERH98(y,2) scaling law [102] (NGE); *middle*: with a correction discounting the "core" radiation (NGE). In either representation, confinement increases even as core emission does too, but this is particularly pronounced when the latter is removed from $\tau_E$. *Top*: estimated inter-ELM loss power fraction through the separatrix (abscissa NGE), which remains constant within uncertainties throughout.

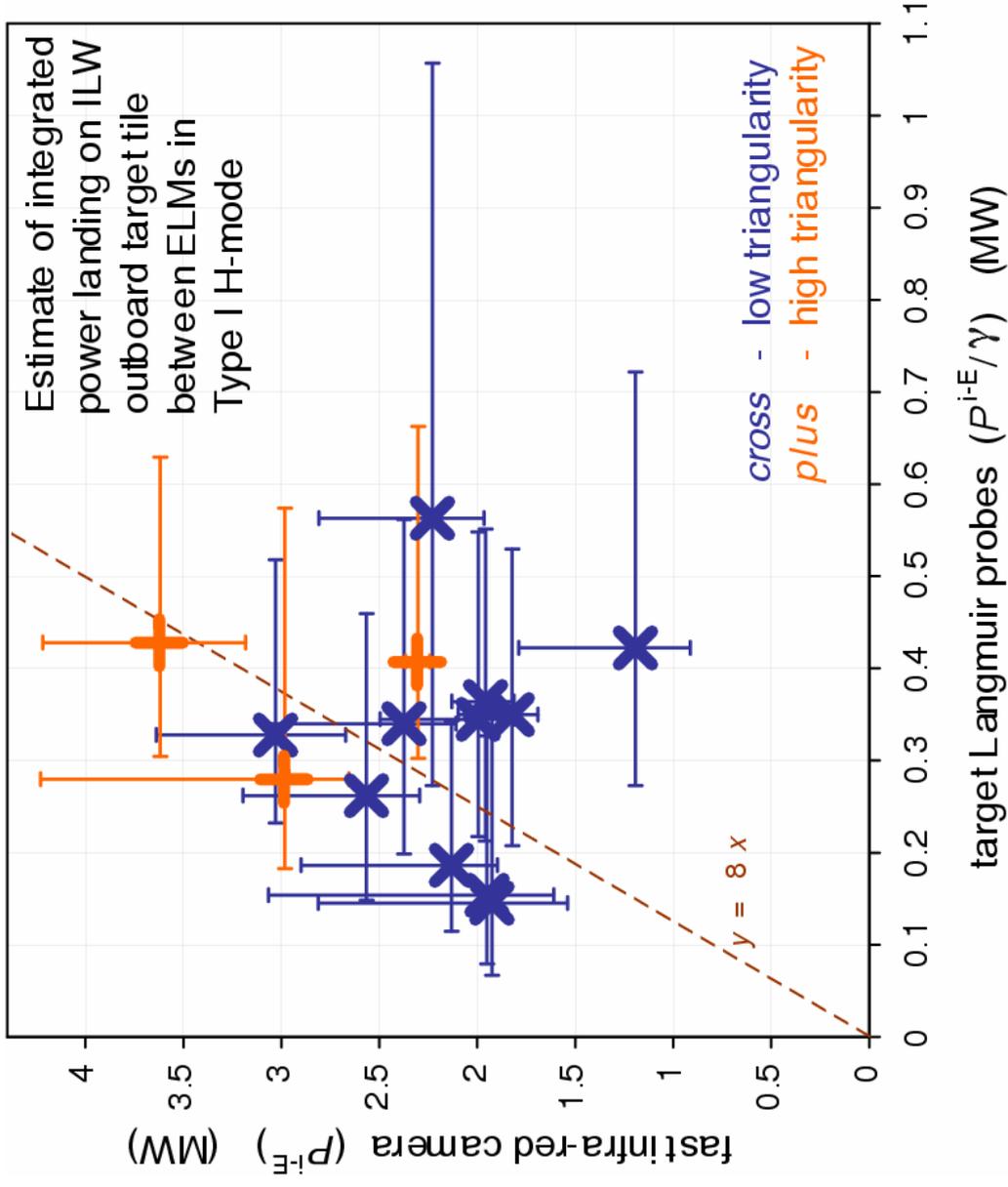

Fig. 9(a) Comparison of estimated total power landing on the divertor whole toroidal outboard target from fast infra-red thermography [133] against integration of embedded Langmuir probes without a sheath transmission factor [145-147], between ELMs during flat-tops of high- (+) and low- (×) triangularity H-modes at 2.5 MA and 16 - 21 MW in ILW fuelling/N-seeding scans. *Superimposed*: exact proportionality with a coefficient of 8. Within significant scatter, Langmuir probe data are approximately consistent with infra-red measurements assuming a sheath total energy transmission factor ($\gamma$) of 8. (NGE)

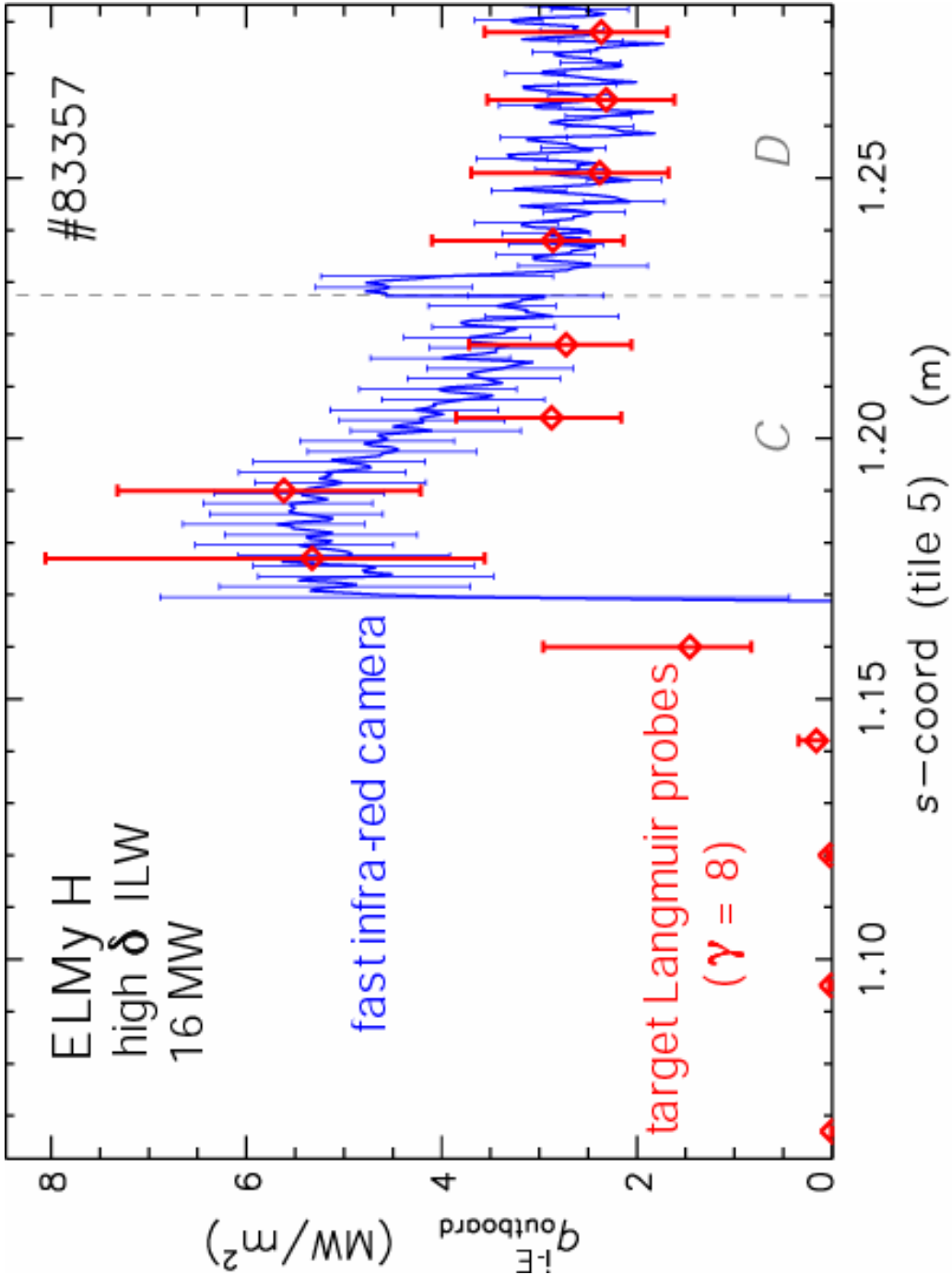

Fig. 9(b) Example comparison of power density across the divertor outboard target from fast infra-red (IR) thermography[133] and embedded Langmuir probes, averaged between ELMs during the flat-top of a high-triangularity, 2.5 MA unseeded H-mode at ≈16 MW in the ILW, against a rectilinear co-ordinate ($s$) along the target surface. The dashed line indicates the boundary between stacks C and D of the plate, at which edge a spike occurs in the IR camera profile. Probe data include a sheath transmission factor of 8 and correction for target inclination, yielding good agreement with thermography. (NGE)
46

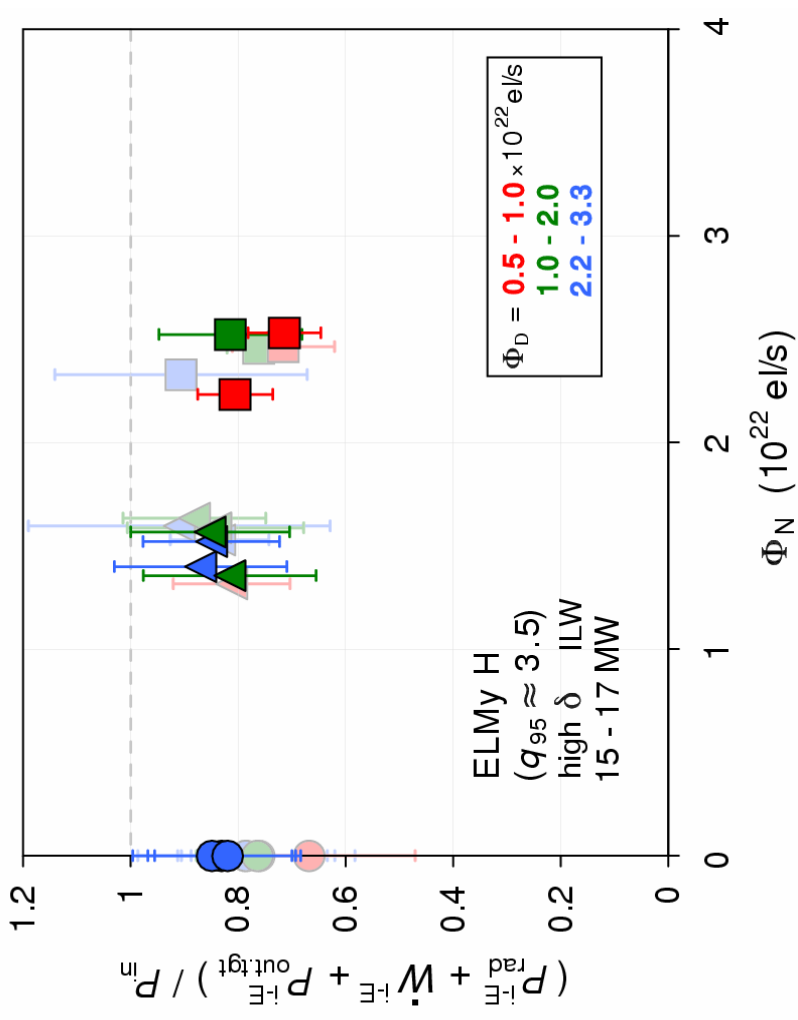
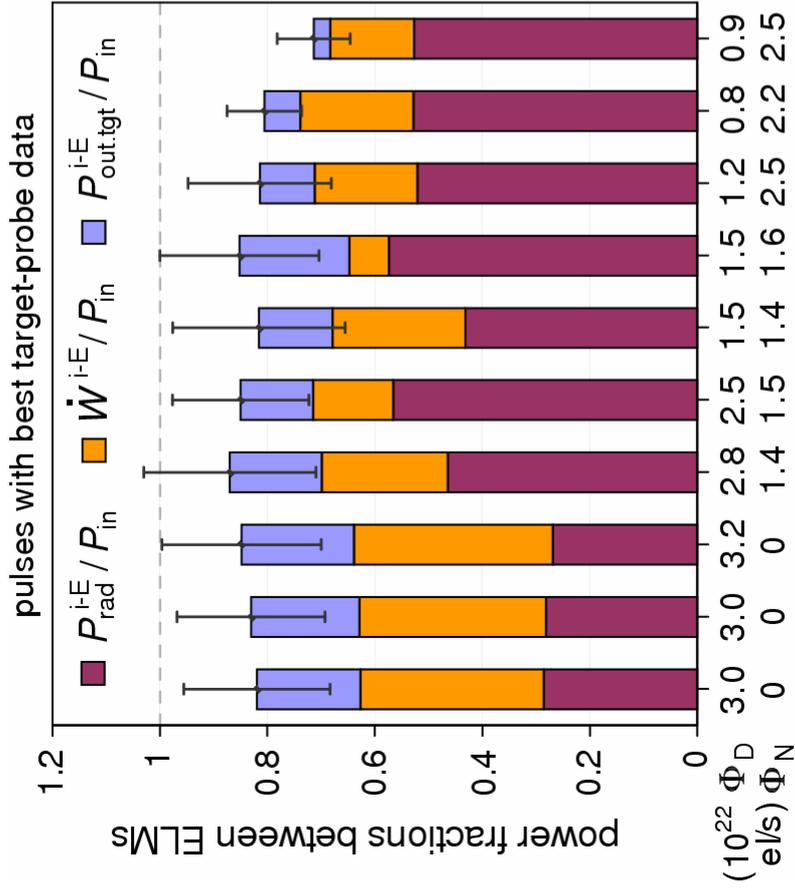

Fig. 10  Analysis of power balance averaged between ELMs during the flat-tops of high-triangularity, 2.5 MA H-modes at 15 - 17 MW in the ILW. *Left*: power fractions radiated, stored by the plasma, or deposited on the divertor outboard target, for those pulses in fuelling and seeding scans with embedded Langmuir probe data ($\gamma = 8$) best defining heat load profiles. *Right*: same net power balance against N-seeding rate (electrons per second assuming full ionisation), points colour and symbol coded for D and N inputs respectively (as in Fig. 3). Pulses with best probe data are shown plain, remaining ones with some significant probes missing are shown faint. Within uncertainties and loss terms (shine-through, inboard target, other surfaces) omitted, a reasonable balance is recovered.

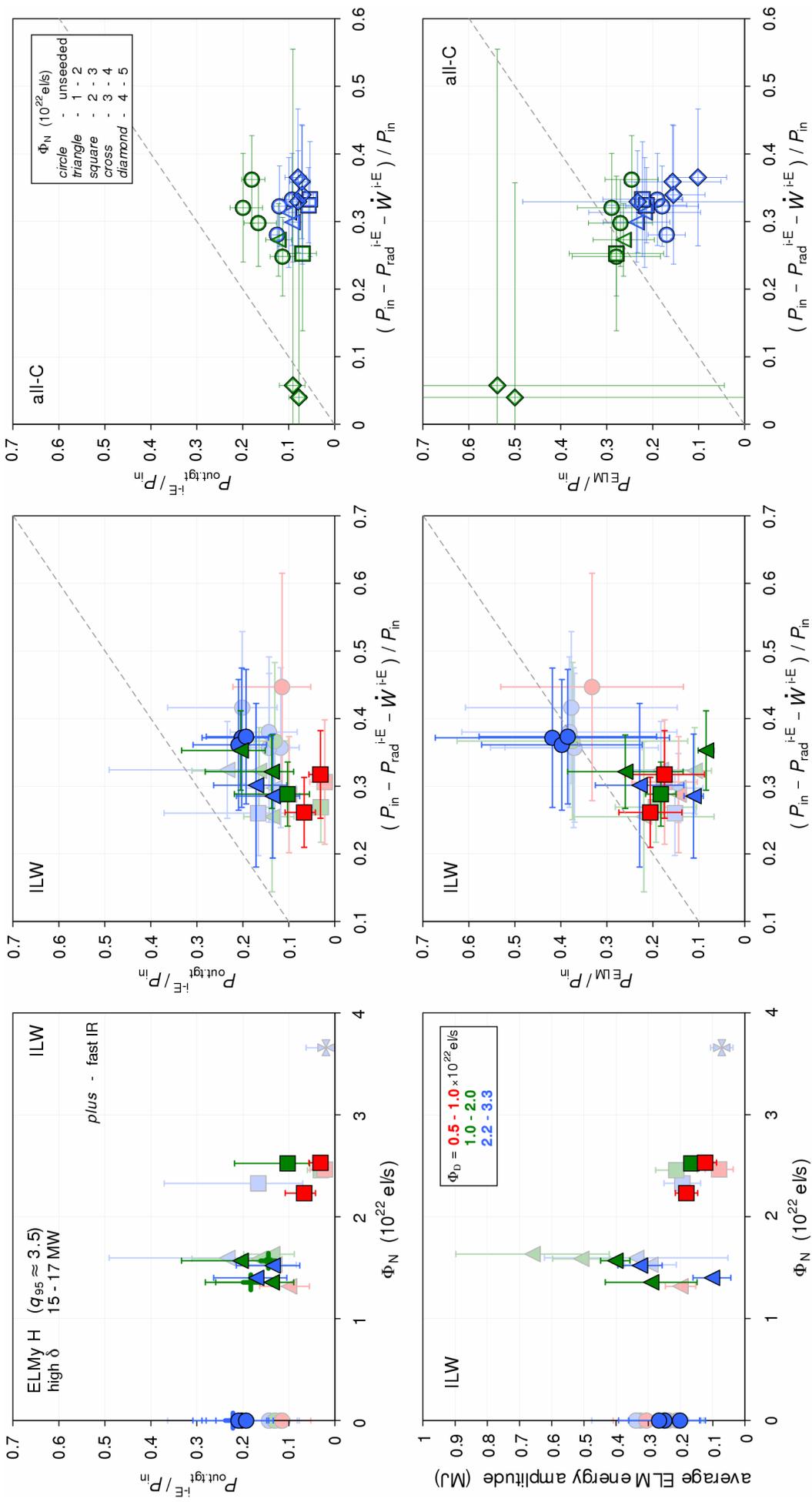

Fig. 11   Divertor outboard target load fraction between ELMs (*top*) and ELM losses (*bottom*) in flat-tops of high-triangularity, 2.5 MA H-modes with power 15 - 17 MW. *Open symbols*: all-C; *filled symbols*: ILW; pulses with best probe data shown plain, remainder shown faint; points colour and symbol coded for D and N rates respectively (as in Fig.3). *Top left*: inter-ELM target load fraction from Langmuir probes versus N input (electrons per second assuming full ionisation) in the ILW, showing a decrease with seeding. (Three corresponding

points (+) from fast IR thermography superimposed). (NGE) *Top centre*: same ordinate versus inter-ELM efflux power fraction, implying the decline in target load is in proportion to the effect of seeding on lowering heat exhaust, except possibly for higher N input at lowest fuelling. (ordinate NGE) *Top right*: equivalent plot for all-C cases, using fast IR thermography for target load between ELMs [42,43]. The changes are smaller, but still roughly in proportion. (ordinate NGE) *Bottom left*: average ELM energy amplitude against N input in the ILW (NB: plain and faint probe sets retained only for easier relation to other plots). The tendency for ELM frequency initially to decrease with seeding leads to reciprocally larger transients. *Bottom centre*: estimated power fraction exhausted in ELMs versus inter-ELM efflux power fraction in the ILW. Surprisingly, the effect of seeding leads to an even stronger relative reduction of ELM power. *Bottom right*: equivalent plot for all-C cases. An opposite trend is suggested, viz. ELM power tends not to change, or even to increase, as heat exhaust is decreased.

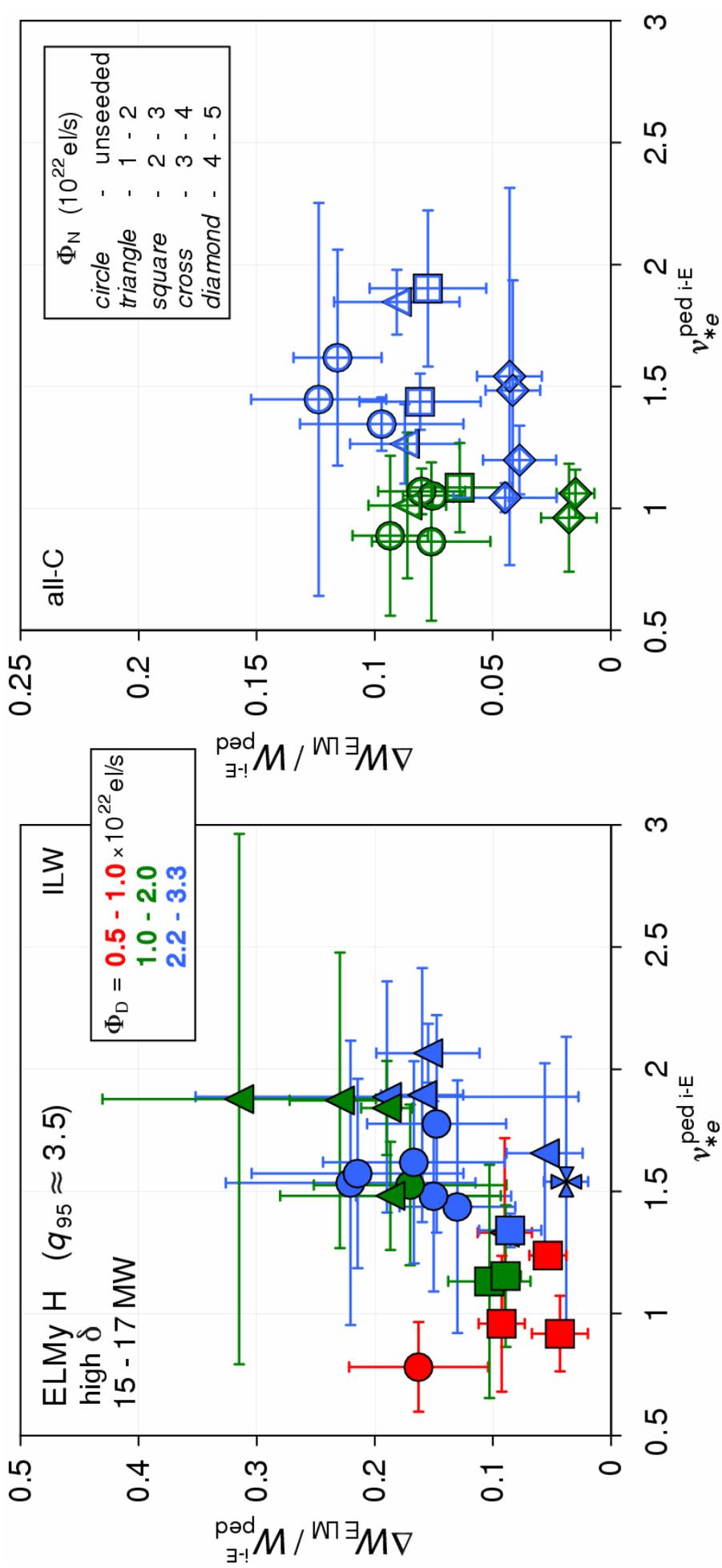

Fig. 12  Average ELM energy amplitude normalised to estimated pedestal energy ($T_i^{\text{ped}} = T_e^{\text{ped}}$), versus electron collisionality[132] at the pedestal top, during flat-tops of high-triangularity, 2.5 MA H-modes with power 15 - 17 MW. Pedestal quantities are inferred from modified tanh fits[126-129] to HRTS profiles ($\rho \geq 0.85$) averaged in the final third of inter-ELM periods (as in Figs. 4(b) & 5). Points colour and symbol coded for D and N inputs respectively (as in Fig. 3). *Left*: ILW; *right*: all-C (note different vertical scale). Although the range in collisionality is narrow, no evidence of the conventional inverse correlation of ELM size[141,36] is seen.

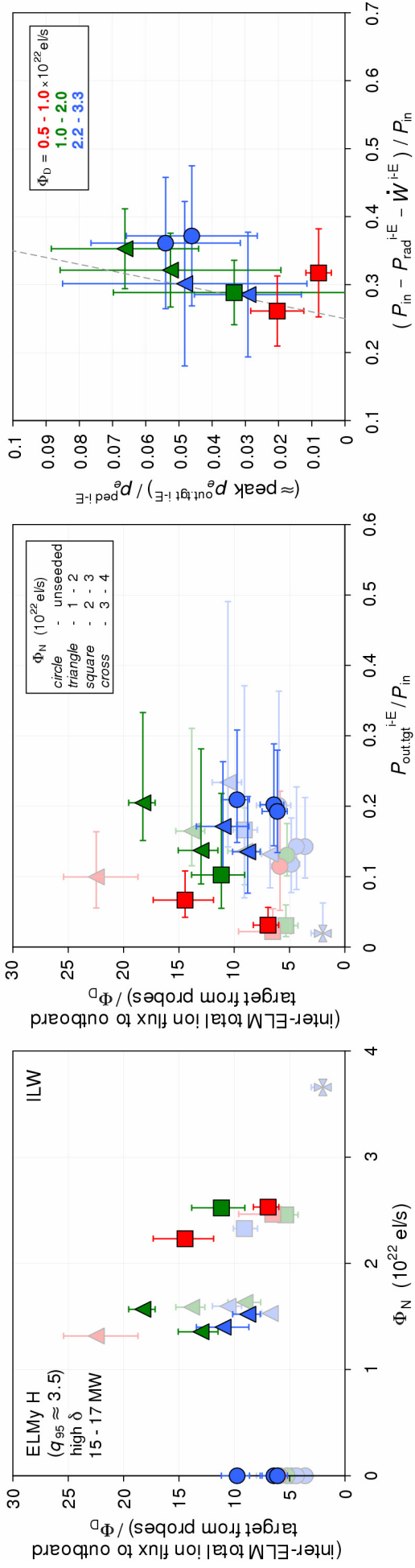

Fig. 13 Inter-ELM outboard detachment during flat-tops of high-triangularity H-modes at 2.5 MA and 15 - 17 MW in the ILW. Points colour and symbol coded for D and N puffs respectively (as in Fig. 3); pulses with best probe data shown plain, others shown faint. *Left*: approximate flux amplification at the divertor outboard target from embedded Langmuir probes versus N-seeding rate (electrons per second assuming full ionisation). A maximum and subsequent decline in flux amplification for rising N input is hinted, particularly at lowest fuelling, suggestive of onset of detachment. (NGE) *Centre*: same ordinate against inter-ELM outboard-target power fraction, also from probes. Flux amplification tends to rise as target load is lowered, with a roll-over tentatively indicated for strongest seeding. (NGE) *Right*: approximate peak electron pressure between ELMs at the outboard target inferred from Langmuir probes, normalised to pedestal pressure derived from modified tanh fits [126–129] to HRTS profiles ($\rho \geq 0.85$) averaged over the final thirds of inter-ELM periods (as in Figs. 4(b) & 5), versus inter-ELM efflux power fraction. Within uncertainties, the estimated decline in target to upstream pressure is in proportion to the cooling effect of seeding.

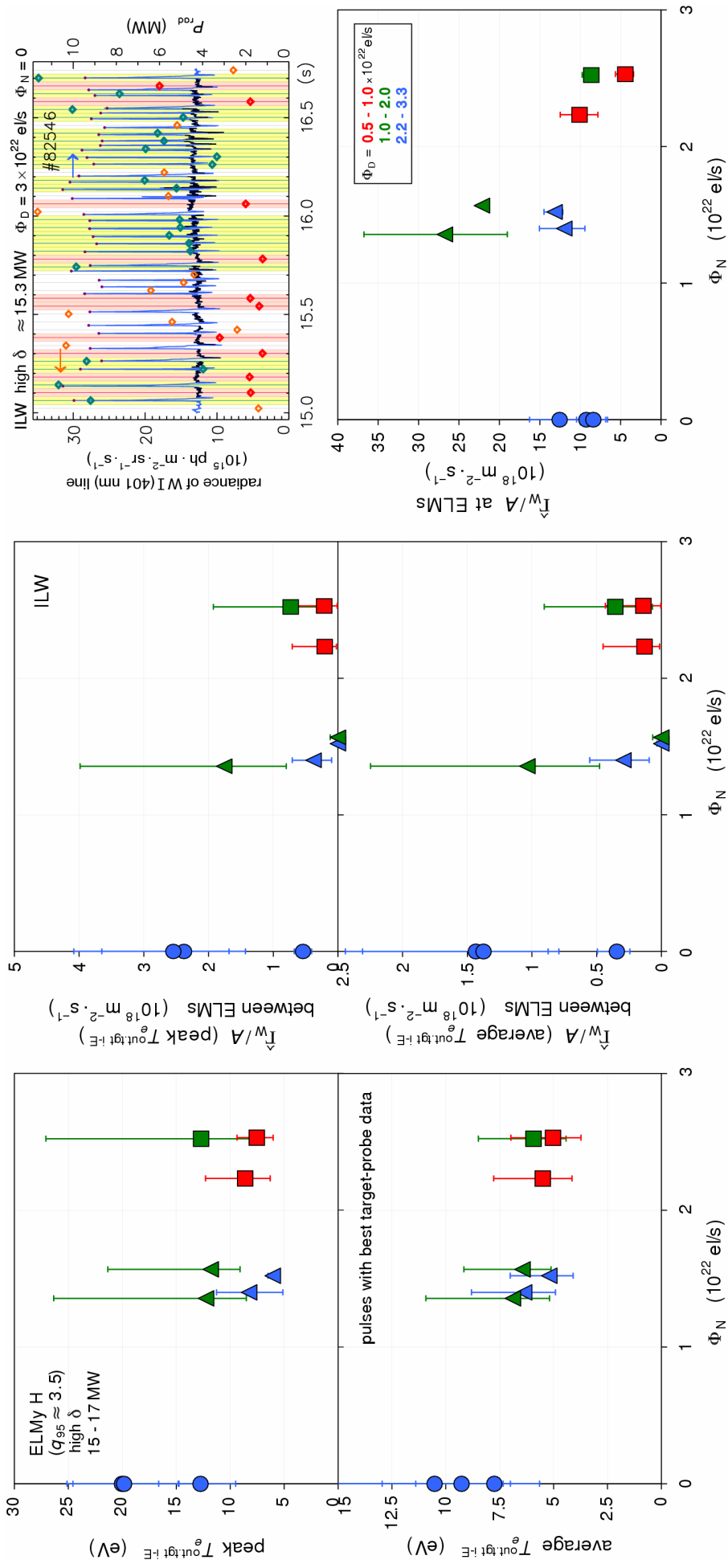

Fig. 14  Electron temperature between ELMs at the divertor outboard target from Langmuir probes (*left*) with estimated tungsten flux per unit area inter-ELM (*centre*) and at ELMs (*bottom right*), as functions of N input (electrons per second assuming full ionisation), during flat-tops of high-triangularity, 2.5 MA H-modes with power 15 - 17 MW and best probe data in the ILW. Points colour and symbol coded for D and N inputs respectively (as in Fig. 3); note the different vertical scales. *Inset top right*: illustration of selecting spectrometer data (*diamonds, left-hand ordinate*) between (*shaded pink*) and at (*shaded yellow*) ELMs identified from total radiated power (*continuous trace, right-hand ordinate*); dark segments are intervals 25% - 90% of the way between peaks. Tungsten flux densities are then derived from averaged radiance of the WI (401 nm) line over the full target width and recent associated $S/(XB)$ data [165,166]. *Left*: peak (*top*) and toroidally-averaged (*bottom*) electron temperature; *centre*: evaluating $S/(XB)$ at these respective (*top, bottom*) figures; *bottom right*: using

asymptotic $S/(XB)$ for high temperatures. (NGE) Near-target electron temperature is moderately reduced by seeding, while the tungsten source is also gradually decreased between ELMs and eventually even modestly at ELMs.

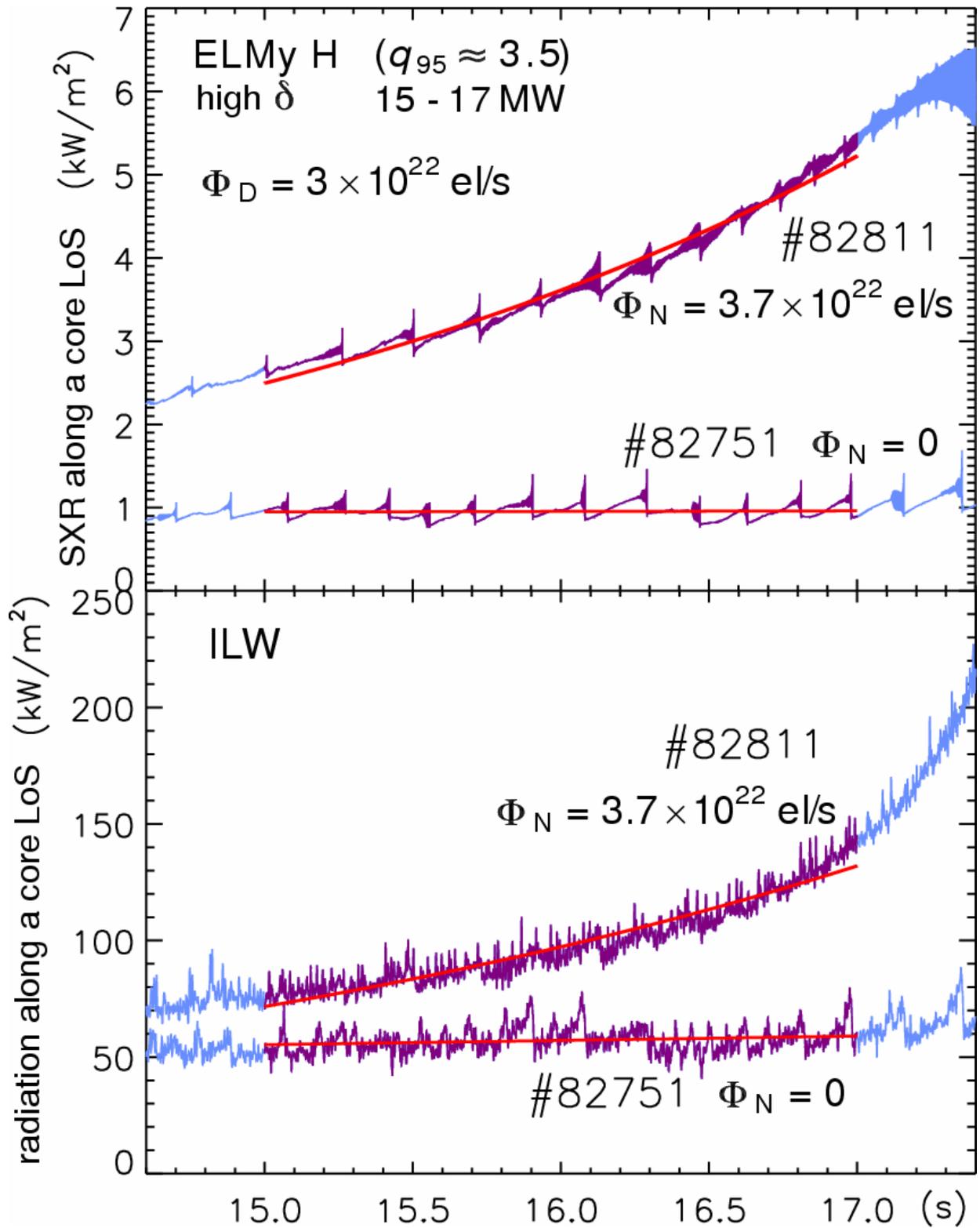

Fig. 15(a)  Example time-traces of soft X-ray (SXR) emission (*top*) and unfiltered radiation (*bottom*) along horizontal lines-of-sight through the confined plasma during the flat-tops of high-triangularity, 2.5 MA, ILW pulses at higher fuelling and unseeded (#82751, ≈ 16 MW) or strongly N-seeded (#82811, ≈ 17 MW). *Superimposed lines*: exponential fits to the shaded segments of each signal. Monotonic increases of both types of emission are evident in the seeded case, while radiation rises even more rapidly after sawteeth become negligible around ≈ 17 s.

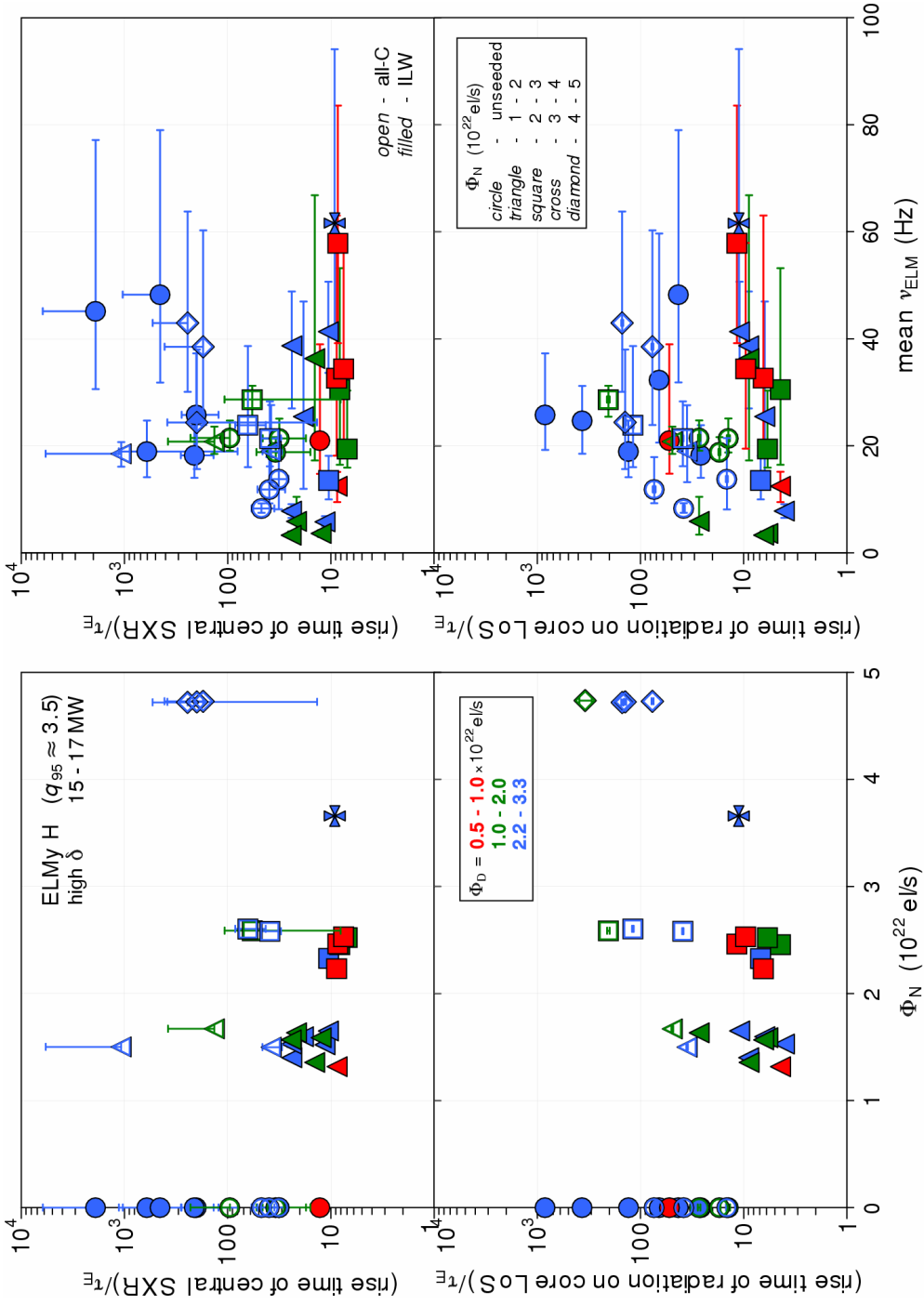

Fig. 15(b) Normalised rise time of SXR (*top*) and unfiltered radiation (*bottom*) along horizontal lines-of-sight through the confined plasma, versus N seeding rate (electrons per second assuming full ionisation) (*left*) and mean ELM frequency (*right*) (abscissa $N_{GE}$), during the flat-tops of high-triangularity, 2.5 MA H-modes at 15 - 17 MW. *Open symbols*: all-C; *filled symbols*: ILW; points colour and symbol coded for D and N inputs respectively (as in Fig. 3). All-C and unseeded ILW plasmas at medium to higher fuelling are generally more steady, but any addition of N in the ILW significantly exacerbates their non-stationarity. There is no indication of a relation to correspondingly altered ELM frequency and amplitude.

55